\documentclass[aps,prb,showpacs,twocolumn]{revtex4-1}
\usepackage{bm,color,amsmath,amssymb,mathrsfs,latexsym,graphicx,psfrag}

\newcommand{\im}{\mathrm{Im}}

\def\ie{{\it i.e.},\ }

\newcommand{\be}{\begin{equation}}
\newcommand{\ee}{\end{equation}}

\def\be{\begin{equation}}
\def\ee{\end{equation}}
\def\bea{\begin{eqnarray}}
\def\eea{\end{eqnarray}}

\def\C60{A$_x$C$_{60}$}

\def\HgCu3{HgCa$_2$Cu$_3$O$_{8+y}$}
\def\HgCu4{HgBa$_2$Ca$_3$Cu$_4$O$_{10+y}$}
\def\TlCu{Tl$_2$Ba$_2$CuO$_{6+\delta}$}
\def\TlCu3{Tl$_2$Ba$_2$Ca$_2$Cu$_3$O$_{10+y}$}
\def\TlCu4{Tl$_2$Ba$_2$Ca$_3$Cu$_4$O$_{12+y}$}

\def\BiCu3{Bi$_2$Sr$_2$Ca$_{2}$Cu$_3$O$_y$}

\def\8LSCO{La$_{1.88}$Sr$_{.12}$CuO$_4$}
\def\110LNSCO{La$_{1.5}$Nd$_{0.4}$Sr$_{0.1}$CuO$_{4}$}
\def\stage4LCO{La$_{2}$CuO$_{4+\delta}$}
\def\Y248{YBa$_2$Cu$_4$O$_8$}

\def\NbSe2{NbSe$_2$}
\def\TaSe2{TaSe$_2$}
\def\TiSe2{TiSe$_2$}

\newcommand{\bk}{{\mathbf k}}

\newcommand{\br}{{\mathbf r}}
\newcommand{\bq}{{\mathbf q}}
\newcommand{\bQ}{{\mathbf Q}}
\newcommand{\Pf}{{\textrm{Pf}}}
\newcommand{\ThetaS}{\hat{\Theta}_{S}}

\begin{document}

\title{Topological Insulators with Commensurate Antiferromagnetism}
\author{Chen Fang$^{1,2}$, Matthew J. Gilbert$^{3,4}$,  B. Andrei Bernevig$^{2}$}
\affiliation{$^1$Department of Physics, University of Illinois, Urbana IL 61801}
\affiliation{$^2$Department of Physics, Princeton University, Princeton NJ 08544}
\affiliation{$^3$Department of Electrical and Computer Engineering, University of Illinois,  Urbana IL 61801, USA}
\affiliation{$^4$Micro and Nanotechnology Laboratory, University of Illinois, Urbana IL 61801, USA}
\date{\today}

\begin{abstract}
We study the topological features of non-interacting insulators subject to an antiferromangetic (AFM) Zeeman field, or AFM insulators, the period of which is commensurate with the lattice period. These insulators can be classified by the presence/absence of an emergent anti-unitary symmetry: the combined operation of time-reversal and a lattice translation by vector $\mathbf{D}$. For AFM insulators that preserve this combined symmetry, regardless of any details in lattice structure or magnetic structure, we show that (i) there is a new type of Kramers' degeneracy protected by the combined symmetry; (ii) a new $Z_2$ index may be defined for 3D AFM insulators, but not for those in lower dimensions and (iii) in 3D AFM insulators with a non-trivial $Z_2$ index, there are odd number of gapless surface modes if and only if the surface termination also preserves the combined symmetry, but the dispersion of surface states becomes highly anisotropic if the AFM propagation vector becomes small compared with the reciprocal lattice vectors. We numerically demonstrate the theory by calculating the spectral weight of the surface states of a 3D TI in the presence of AFM fields with different propagation vectors, which may be observed by ARPES in Bi$_2$Se$_3$ or Bi$_2$Te$_3$ with induced antiferromagnetism.
\end{abstract}
\maketitle

\section{Introduction}

The inception point in the study of topological phases in condensed matter systems is the integer quantum Hall effect (IQHE). Since the Hall conductivity is odd under time-reversal symmetry (TRS), the topologically non-trivial states will only occur when TRS is broken. The connection between the observed quantized Hall conductance and topology is made through the calculation of the Chern number which, in the context of the IQHE, corresponds to the number of chiral edge states present and is a global quantity contributed by all of the occupied states in the system\cite{Thouless1982}. Nevertheless, the vast majority of topological band insulator states that have been experimentally confirmed\cite{MarkusKonig11022007,Hsieh:2008fk,roushan2009,Checkelsky2011}, are stabilized by the \emph{presence} of TRS in which strong spin-orbit interactions invert the bands which allows the creation of a mass boundary \cite{Haldane1988,Hasan2010,qi2011rev} that traps surface or edge states at the interface of two bulk band insulators or that of a bulk insulator and vacuum, commonly referred to as topological insulators (TI). While the theory associated with TRS TI is well understood, the presence of TRS in a material significantly constrains the number of potential topological materials. Therefore, it is important to go beyond the well-established paradigm of non-interacting time-reversal symmetric TI to look for novel topological phases. To do so, one must generalize the theory of topological classification to encompass a wider spectrum of materials\cite{Schnyder2008,Kitaev2009,Ryu2010a,hughes2010inv,Turner:2012,Alexandradinata2012,Fang2012,Fang:2011,Fu:2012,Xu2012,Tanaka2012}. 

Of the materials not covered in the standard theory of TRS TI, one of the most interesting classes are magnetic materials. The established theory for TRS insulators cannot be applied to classify any magnetic materials as their ground states break TRS. Ferromagnetically ordered materials have been explored theoretically as their broken symmetry leads to the the possible condensed matter realization of Weyl fermions\cite{Wan2011,Fang:2011,XuG2011}. (While the magnetic structure studied in Ref.[\onlinecite{Wan2011}] has zero total magnetization, we still refer to it as ferromagnetism because of the unbroken translational symmetry.) Yet, to this point, there has been comparatively little work on antiferromagnetically ordered materials. A pioneering work by Mong, Essin and Moore\cite{Mong2010} showed that a new $Z_2$-index can be defined in an antiferromagnet where the unit cell is doubled by the magnetic structure, such that while both time-reversal and translation along a primitive basis vector are broken, their combination is preserved. The spin structure in these insulators can be seen as sets of ferromagnetically ordered 2D planes of spins with alternating magnetization, stacked along a crystal direction $[mnl]$. Nevertheless, real magnetic materials usually exhibit more complex magnetic structures, e.g., having non-collinear configurations as spiral magnetism, having a very large magnetic unit cell (small propagation vectors) or having more than one propagation vector. These possibilities are specially relevant for AFM field induced by dopants in an otherwise non-magnetic insulator like Bi$_2$Se$_3$ or Bi$_2$Te$_3$. Therefore, it is necessary to build a $Z_2$ classification for AFM insulators with generic magnetic ordering, which we do in this paper.

Let us recall that the existence of Kramers' degeneracy\cite{Kramers1930} is necessary for defining the $Z_2$ index for TRS insulators\cite{kane2005B,Bernevig:2006kx,Fu:2006rm,Fu:2007fk,zhang2009}. Kramers' degeneracy is the exact double degeneracy of every single particle state in a spin-$1/2$ system with TRS. We begin by studying the breaking or preservation of Kramers' degeneracy in AFM insulators. While it is clear that ferromagnetism always breaks a Kramers' pair into energetically separated spin-up and spin-down states, we show that AFM preserves the degeneracy of almost all single particle states, with exceptions detailed in later sections, if there is a lattice translation vector $\mathbf{D}$ that inverts the spin at every site. This is because if such $\mathbf{D}$ exists, we can define an anti-unitary operator\cite{Mong2010} $\Theta_S=\Theta{T}_{\mathbf{D}}$, where $\Theta$ is the time-reversal symmetry operator, and $T_{\mathbf{D}}$ represents a translation by $\mathbf{D}$. $\Theta_S$ is a symmetry because both $\Theta$ and $T_{\mathbf{D}}$ invert the magnetization and so their combination recovers it. It is this symmetry that ensures that Kramers' degeneracies are preserved for almost all single particle states. AFM insulators that are invariant under $\Theta_S$ will be referred to as $\Theta_S$-symmetric insulators hereafter.

We may arrive at the same result beginning from a different perspective, by treating the AFM field, a static field coherently coupled to the electrons in the form of $c^\dag_{\tau}(\br)(\mathbf{J}(\br)\cdot\mathbf{\sigma})_{\tau\tau'}c_{\tau'}(\br)$, where $\mathbf{\sigma}$ is the spin operator and $\tau,\tau'$ the spin indices, as perturbation and expressing the single-particle Green's function to an arbitrary order of perturbation. This allows us to calculate the poles of the Green's function to obtain the energy spectrum. From this perturbative point of view, the energy spectrum of the TRS insulator is modified by the self-energy contribution from scattering by the AFM field, and Kramers' degeneracy would require that the self-energy term be invariant under TRS. The full self-energy counts in all scattering processes that send the particle to its initial state. Since the AFM field changes sign under TRS, the scattering contribution from even number of scatterings are invariant under TRS and those from odd number of scatterings are variant. Therefore, the total self-energy term is invariant under TRS if and only if there are only even number of scatterings. Any commensurate AFM field decomposes into a finite series of normal modes, namely $\mathbf{J}(\br)=\sum_{i=1,...,m}\mathbf{M}(\bQ_i)e^{i\bQ_i\cdot\br}$, where $\bQ_{i=1,...,m}$ are the set of propagation vectors of the AFM field. Consider an electron with initial momentum $\bk$, then after $N$ scatterings its momentum becomes $\bk'=\bk+z_1\bQ_1+z_2\bQ_2+...+z_m\bQ_m$, where $z_1+z_2+...+z_m=N$. If this scattering process contributes to the self-energy, we have $\bk'=\bk+\mathbf{G}$, where $\mathbf{G}$ is any reciprocal lattice vector. Preservation of the Kramers' degeneracy thus requires that only $N=even$ terms are nonzero. From this, we obtain the necessary and sufficient condition for existence of a Kramers' degeneracy: for any set of $m$ integers $(z_1,...,z_m)$ satisfying $\sum_{i}z_i\bQ_i=\mathbf{G}$, there is $\sum_iz_i=even$. Mathematically, satisfying this condition is completely equivalent to the existence of a lattice translation vector $\mathbf{D}$ that inverts the AFM field. This explicitly indicates that the propagation vectors alone determine whether a Kramers' degeneracy of $\Theta_S$ is preserved or broken, while any other detail in the magnetic structure is irrelevant. Also, we use the perturbation theory to estimate the split of the Kramers' pair when the above condition is violated.

Having established the necessary and sufficient conditions for the Kramers' degeneracy in a commensurate AFM, we further investigate if this allows us to define a $Z_2$ topological invariant. It is known that one can define the $Z_2$ invariant using the Pfaffians of the sewing matrix of time-reversal operator at time-reversal invariant momenta (TRIM) in 2D and 3D TRS insulators\cite{kane2005B,Fu:2006rm,Fu:2007fk}. We prove that in $\Theta_S$-symmetric AFM insulators, a $Z_2$-invariant can be defined only in 3D (or certain higher dimensions) using Pfaffians of the sewing matrix associated with $\Theta_S$ at half of the TRIM, while in 2D and 1D, the same definition gives a gauge variant quantity. This is because the $d$-dimensional AFM $Z_2$ number is defined in its $d-1$-dimensional subsystem that belongs to the symplectic class (AII in the A-Z classification\cite{Altland1997} of all Hamiltonians), and 0D and 1D symplectic Hamiltonians are always topologically trivial\cite{Schnyder2008,Kitaev2009,Ryu2010a}. However, when in addition to $\Theta_S$ there is also spatial inversion symmetry, one can use inversion eigenvalues at half of the occupied bands to define a $Z_2$-invariant in arbitrary dimensions.

In TRS insulators there is a correspondence between the bulk $Z_2$-invariant and the existence of gapless boundary modes\cite{qi2006}: when the $Z_2$ number is non-trivial, there are odd number of gapless (Dirac) modes at the boundary of the system. Here we show that the existence of gapless surface states in a 3D AFM insulator requires two conditions: (i) it has a non-trivial $Z_2$-invariant in the bulk and (ii) the surface termination \emph{also preserves} the $\Theta_S$-symmetry, i.e., the $\Theta_S$ symmetry is unbroken on the surface. If the AFM is small, one may say that the Dirac surface modes of 3D TRS TI are preserved in presence of a $\Theta_S$-symmetric AFM. However, in a 2D $\Theta_S$-symmetric AFM insulator, the surface modes are \emph{not} protected from being pushed into the bulk by tuning the surface chemical potential. This is in exact agreement with the fact that the $Z_2$-invariant is only well-defined in 3D $\Theta_S$-symmetric AFM insulators and not in lower dimensions.

While a TI that has intrinsic AFM ordering is yet to be experimentally established, the theory can be applied to TI thin films with induced AFM ordering by an AFM substrate. In principle one can use the substrates to induce various AFM orderings in the TI thin film, but lattice matching at the interface is a practical issue. The lattice constants of the substrate should be nearly identical to those of the TI at the interface, or at least commensurate with them. If the AFM substrate is $\Theta_S$-symmetric in the redefined basis, the Dirac point on the surface of the TI would be unbroken by the induced AFM, while the surface state dispersion becomes anisotropic along the directions parallel and perpendicular the AFM propagation vector.

It is straightforward to extend the discussion to topological superconductors (TSC) with coexisting AFM orders. Given a $d$-dimensional TSC in the DIII class (SC's with non-negligible spin orbital coupling and with TRS), which has an induced or intrinsic AFM order, if the AFM ordering is $\Theta_S$-symmetric, then we show that the AFM TSC has the same classification as a $d-1$-dimensional TSC without AFM. Therefore AFM SC's in 3D and 2D have $Z_2$ topological classifications\cite{Schnyder2008,Qi2010,Ryu2010a}.

The paper is organized as follows. In Sec.\ref{sec:Kramer}, we show that a new type of degeneracy is preserved in an AFM insulator if and only if it is $\Theta_S$-symmetric. In Sec.\ref{sec:Perturbation}, we use perturbation expansion of self-energy in the single particle Green's function to re-derive the condition for Kramers' degeneracy, and, when the condition is not met, the relation between the energy splitting of a Kramers' degeneracy by the AFM and the propagation vectors of the AFM. In Sec.\ref{sec:Z2}, the results obtained in the previous sections are used to classify 3D AFM insulators by a $Z_2$-invariant. In Sec.\ref{sec:Boundary} we study the boundary modes of 2D and 3D AFM insulators, reaching the conclusion that only 3D $\Theta_S$-symmetric AFM insulators may have protected gapless Dirac modes, yet anisotropic, on the surface that is also $\Theta_S$-symmetric. In Sec.\ref{sec:discussion}, we discuss how the theory may be applied in real materials and also how an analysis similar to what we perform in AFM insulators can be easily extended to AFM superconductors. We conclude the work in Sec.\ref{sec:Conclusion}.

\section{Kramers' degeneracy in an AFM insulator: proof by symmetries}\label{sec:Kramer}

Throughout the paper, we assume that the electrons in the AFM insulator can be modeled by a TRS tight-binding Hamiltonian that coherently couple to an antiferromagnetic Zeeman field:
\bea\label{eq:first_eq}
\hat{H}=\hat{H}_0+\hat{H}_M,
\eea 
where
\bea
\hat{H}_0=\sum_{\bk,\alpha,\beta,\tau,\tau'}\mathcal{H}_0^{\alpha\tau,\beta\tau'}(\bk)c^\dag_{\alpha\tau}(\bk)c_{\beta\tau'}(\bk)
\eea 
is the hopping part that is TRS and
\bea
\hat{H}_M=\sum_{\br,\alpha,\tau,\tau'}(\mathbf{J}(\br)\cdot\sigma_{\tau\tau'})c^\dag_{\alpha\tau}(\br)c_{\alpha\tau'}(\br)
\eea
is the part coupled to the Zeeman field $\mathbf{J}(\br)$. In the above expressions and hereafter, Greek letters denote the orbital degrees of freedom inside a unit cell with exceptions of $\tau,\tau'$ which refer to up and down spin components. We assume $\mathbf{a}_{i=1,...,d}$'s to be the basis vectors of the $d$-dimensional lattice in which the model is embedded; and $\mathbf{b}_{i=1,...,d}$'s are the corresponding reciprocal lattice basis vectors. The AFM field $\mathbf{J}(\br)$ breaks the lattice translation symmetry, but as long as its periods are commensurate with lattice vectors, it enlarges the unit cell of the crystal lattice. The enlarged unit cell, or magnetic unit cell, is given by the basis vectors $\mathbf{a}^M_{i}$'s, which is in general a linear combination of the $\mathbf{a}_i$'s with integer coefficients. The counterpart of the enlarged unit cell in the reciprocal space is the folded Brillouin zone (BZ), or the magnetic BZ (MBZ) spanned by basis vectors $\mathbf{b}^M_i$'s satisfying $\mathbf{a}^M_i\cdot\mathbf{b}^M_j=2\pi\delta_{ij}$.

Before proceeding to discuss the degeneracy in a time-reversal breaking AFM system, it is helpful to briefly review the concept of Kramers' degeneracy in a TRS system, and the key role it played in the $Z_2$-classification of TRS insulators in 2D and 3D. The time-reversal operator in the single-fermion sector of Hilbert space, denoted by $\hat{\Theta}$ is defined as:
\bea\label{eq:Theta1}\hat{\Theta}&=&K(i\sigma_y),
\eea 
where $K$ is the complex conjugate operator. From this definition one can see that $\hat\Theta$ is antiunitary and squares to minus the identity. These two properties guarantee that each single particle state is at least doubly degenerate\cite{Kramers1930}. This degeneracy protected by time-reversal symmetry is called Kramers' degeneracy and the degenerate doublet referred to as a Kramers' pair. In a translationally invariant system, one can further show that an eigenstate with momentum $\bk$ is sent by $\hat\Theta$ to an equal energy eigenstate at momentum $-\bk$. Specially, at center or corner of BZ where we have $-\bk_{inv}=\bk_{inv}\;\textrm{mod}\;\mathbf{G}$ hence every energy level of $\hat{H}(\bk_{inv})$ must be doubly degenerate. These points at BZ center and corners are called time-reversal invariant momenta (TRIM) and the Kramers' degeneracy at TRIM make possible the definition of a $Z_2$-invariant in 2D and 3D TRS insulators. In order to formally define this invariant, one uses the sewing matrix, which is defined as
\bea
\mathcal{B}_{mn}(\bk)=\langle\psi_m(-\bk)|\hat{\Theta}|\psi_n(\bk)\rangle,
\eea 
where $|\psi_m(\bk)\rangle$ is the energy eigenstate at $\bk$ in the $m$-th band. The sewing matrix is antisymmetric at all TRIM and this antisymmetry enables the definition of the Pfaffians at TRIM which form the basis of the definition of the $Z_2$-invariant $\delta_0$, given by (Fu-Kane formula\cite{Fu:2007fk}):
\bea
(-1)^{\delta_0}=\prod_{\bk_{inv}}\frac{\Pf[\mathcal{B}(\bk_{inv})]}{\sqrt{\det[\mathcal{B}(\bk_{inv})]}}.
\eea 
It should be noted that although it appears that $\delta_0$ only depends on the band structure at TRIM, it implicitly depends on the band structure in the whole BZ, as a smooth gauge is required for this definition.

With the role of Kramers' degeneracy defined in TRS insulators, let us turn our attention to the AFM insulators. Any AFM breaks TRS because TRS reverses all spins leaving the orbital and spatial components invariant as:
\bea\label{eq:Theta3}
\hat{\Theta}\hat{H}_M\hat\Theta^{-1}=-\hat{H}_M.
\eea 
Therefore, Kramers' degeneracy in its original meaning does not exist in any AFM insulator. However, this does not necessarily mean that the spectrum is non-degenerate. In fact, each energy level may still be doubly degenerate, but the two states are not each other's time-reversed counterparts. In that case, there must be some other symmetry that protects the pair from splitting in energy. For a generic AFM TI, the only symmetry is the magnetic translation symmetry by any magnetic superlattice vector $\mathbf{L}^M$ (a superposition of $\mathbf{a}^M_i$'s with integer coefficients), i.e.,
\bea\label{eq:temp11}
[\hat{H},\hat{T}_{\mathbf{L}^M}]=0,
\eea 
where $\hat{T}_{\mathbf{L}}$ is the operator for a translation by a lattice vector $\mathbf{L}$. Eq.(\ref{eq:temp11}) states that one can find a set of basis vectors that are common eigenstates of $\hat{H}$ and $\hat{T}_{\mathbf{L}^M}$. The definition of $\hat{T}_{\mathbf{L}^M}$ dictates that its eigenvalues must be of the form $\exp(i\bk\cdot\mathbf{L}^M)$, where $\bk$ is constrained within the first MBZ. This symmetry gives us a band structure defined in MBZ, but does not protect any degeneracy.

\begin{figure}[!htb]
\includegraphics[width=8cm]{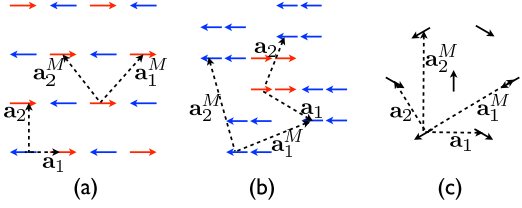}
\caption{AFM configurations that are (a,b) $\Theta_S$-invariant and (c) $\Theta_S$-variant. In (a), we plot the typical spin structure of a CuO$_2$ layer in any parent compound of cuprates superconductors\cite{Tsuei2000}. In (b), we show the spin structure of vacancy doped iron-based superconductor Rb$_2$Fe$_4$Se$_5$\cite{Wang2011}. In (c), we plot the ground state of a classical Heisenberg model on a triangular lattice. The basis vectors of the lattice, $\mathbf{a}_{1,2}$, and those of the magnetic superlattice, $\mathbf{a}^M_{1,2}$ are also shown. Note that in (a) and (b), a translation by $\mathbf{D}=\mathbf{a}_1$ inverts all spins, while in (c) one cannot find such a vector.}
\label{fig:structure}
\end{figure}

In a large class of magnetic structures, examples of which are shown in Fig.\ref{fig:structure}(a,b), there exists a special lattice vector $\mathbf{D}$, where, after a translation by $\mathbf{D}$, all spins flip their signs. Mathematically, we may define this as
\bea\label{eq:D}\{\hat{T}_{\mathbf{D}},\hat{H}_M\}=0.\eea The translation vector, $\mathbf{D}$, is unique only up to some magnetic translation vector $\mathbf{L}^M$. We now suppose that both $\mathbf{D}_1$ and $\mathbf{D}_2$ satisfy Eq.(\ref{eq:D}), then we have
\bea[\hat{T}_{\mathbf{D}_1-\mathbf{D}_2},\hat{H}_M]=[\hat{T}_{\mathbf{D}_1}\hat{T}^{-1}_{\mathbf{D}_2},\hat{H}_M]=0,\eea which indicates $\mathbf{D}_1-\mathbf{D}_2=\mathbf{L}^M$. Using the translational invariance and TRS of the free part $\hat{H}_0$, we have
\bea\label{eq:temp1}[\hat{H}_0,\hat{T}_\mathbf{D}]=[\hat{H}_0,\hat\Theta]=0;\eea using Eq.(\ref{eq:Theta3}) in conjunction with Eq.(\ref{eq:D}) we have
\bea
\label{eq:temp2}[\hat{H}_M,\hat{T}_\mathbf{D}\hat{\Theta}]=0.
\eea 
Combining Eq.(\ref{eq:temp1}) and Eq.(\ref{eq:temp2}), we obtain a new symmetry of the full Hamiltonian, $\hat{\Theta}_S=\hat{\Theta}\hat{T}_{\mathbf{D}}$:
\bea\label{eq:ThetaS1}[\hat{H},\hat{\Theta}_S]=0.\eea Since $\hat{T}_\mathbf{D}$ is unitary and $\hat\Theta$ antiunitary, $\hat\Theta_S$ is antiunitary, namely:
\bea\label{eq:ThetaS2}\langle\psi|\phi\rangle=\langle\hat{\Theta}_S\phi|\hat{\Theta}_S\psi\rangle.\eea On the other hand square of $\hat{\Theta}_S$ is
\bea\label{eq:ThetaS3}\hat{\Theta}_S^2=\hat\Theta^2\hat{T}_{2\mathbf{D}}=-\hat{T}_{2\mathbf{D}},\eea where we have used the fact that $\hat{\Theta}$ and $\hat{T}_\br$ commute.

We now turn our attention to the degeneracy in the spectrum of $\hat{H}$. Suppose $|\bk,n\rangle$ is an eigenstate of both $\hat{H}$ and $\hat{T}_{\mathbf{L}^M}$ with $\bk$ a wave vector in the MBZ, then
\bea\hat{H}|\bk,n\rangle&=&E_n(\bk)|\bk,n\rangle\\
\hat{T}_{\mathbf{L}^M}|\bk,n\rangle&=&\exp(i\bk\cdot\mathbf{L}^M)|\bk,n\rangle,\eea then from Eq.(\ref{eq:ThetaS1}) we have
\bea\hat{H}\hat{\Theta}_S|\bk,n\rangle=E_n(\bk)\hat{\Theta}_S|\bk,n\rangle,\eea meaning that $\ThetaS|\bk,n\rangle$ is also an eigenstate of $\hat{H}$ with the same eigenvalue. To check if $|\bk,n\rangle$ and $\ThetaS|\bk,n\rangle$ are same or different state, we calculate the overlap using Eq.(\ref{eq:ThetaS2}) and Eq.(\ref{eq:ThetaS3}) as
\bea\label{eq:temp3}\langle\bk,n|\ThetaS|\bk,n\rangle&=&\langle\ThetaS^2\bk,n|\ThetaS|\bk,n\rangle\\
\nonumber&=&-\langle\bk,n|\hat{T}_{-2\mathbf{D}}\ThetaS|\bk,n\rangle.\eea At this point we notice that $2\mathbf{D}$ is a magnetic superlattice vector as it leaves $\hat{H}_M$ invariant, which implies that
\bea\label{eq:temp4}\langle\bk,n|\hat{T}_{-2\mathbf{D}}=\langle\bk,n|e^{-2i\bk\cdot\mathbf{D}}.\eea
Substitute Eq.(\ref{eq:temp4}) into Eq.(\ref{eq:temp3}) and have
\bea
\label{eq:overlap}\nonumber\langle\bk,n|\ThetaS|\bk,n\rangle=-e^{-2i\bk\cdot\mathbf{D}}\langle\bk,n|\ThetaS|\bk,n\rangle.\\\eea
Eq.(\ref{eq:overlap}) shows that unless $2\bk\cdot\mathbf{D}=(2n+1)\pi$, we have $\langle\bk,n|\ThetaS|\bk,n\rangle=0$, ensuring that $|\bk,n\rangle$ and $\ThetaS|\bk,n\rangle$ are two orthogonal states. Beyond this, it is crucial to understand how the state $\ThetaS|\bk,n\rangle$ transforms under a superlattice translation or
\bea
\label{eq:translation}\hat{T}_{\mathbf{L}^M}\ThetaS|\bk,n\rangle&=&\ThetaS\hat{T}_{\mathbf{L}^M}|\bk,n\rangle\\
\nonumber&=&\ThetaS{e}^{i\bk\cdot\mathbf{L}^M}|\bk,n\rangle\\
\nonumber&=&e^{-i\bk\cdot\mathbf{L}^M}\ThetaS|\bk,n\rangle.\eea Therefore $\ThetaS|\bk,n\rangle$ must be proportional to a state with wavevector $-\bk$, or mathematically
\bea|-\bk,n\rangle=\sum_{n\in{occ}}\mathcal{B}_{mn}(\bk)|\bk,n\rangle,\eea where $m,n$ are band indices ($m\neq{n}$ in general due to degeneracy) and $\mathcal{B}_{mn}(\bk)$ is called the sewing matrix of $\Theta_S$ at $\bk$. Moreover, from Eq.(\ref{eq:translation}), we also notice that other than at TRIM $\ThetaS|\bk,n\rangle$ is orthogonal to $|\bk,n\rangle$ for a generic $\bk$ because they have different eigenvalues under superlattice translation. 

To this point, we have shown that at all non-TRIM, the eigenstate at $\bk$ must be degenerate with another state at $-\bk$. Furthermore, at TRIM for which $\exp(i2\bk_{inv}\cdot\mathbf{D})=1$, each level at $\bk_{inv}$ must be doubly degenerate and that at TRIM for which $\exp(i2\bk_{inv}\cdot\mathbf{D})=-1$, the levels at $\bk_{inv}$ are generically \emph{non-degenerate} if there is no other symmetry present in the system. As we know that the TRIM with degenerate energy levels are important for the definition of $Z_2$-invariant in a TRS insulator, thus it is important to distinguish the TRIM with $\exp(i2\bk_{inv}\cdot\mathbf{D})=-1$. Since $2\mathbf{D}$ is a superlattice vector, we have the general form of $\mathbf{D}$ as
\bea\label{eq:generalD}
\mathbf{D}=\sum_{i=1,...,d}x_i\mathbf{a}^M_i/2+\mathbf{L}^M,
\eea 
where $x_i$ is either zero or one while there is at least one  $x_i=1$. On the other hand, all TRIM can be represented by
\bea\label{eq:ATRIM}
\bk_{inv}=\sum_{i=1,...,d}y_i\mathbf{b}^M_i/2,
\eea 
where $y_i$ is either zero or one. Then if $\exp(i2\bk_{inv}\cdot\mathbf{D})=-1$, we have, using $\mathbf{a}^M_i\cdot\mathbf{b}^M_j=2\pi\delta_{ij}$,
\bea\label{eq:temp5}\sum_{i=1,...,d}x_iy_i\in{odd}.\eea 
There are $2^d$ TRIM, and for a given set of $\{x_i\}$, exactly half of them satisfy Eq.(\ref{eq:temp5}). The proof goes as follows. Since there must be one $x_i$ that is nonzero, we can assume $x_1=1$ without loss of generality. Then for any set of $\{y_1,y_2,...,y_d\}$ that satisfies Eq.(\ref{eq:temp5}), $\{1-y_1,y_2,...,y_d\}$ satisfies $\sum_ix_iy_i=even$. So there is a one-to-one correspondence between TRIM that satisfy Eq.(\ref{eq:temp5}) (called from now on A-TRIM) and those that do not (called B-TRIM). At an A-TRIM, each level is non-degenerate while at a B-TRIM, each level is doubly-degenerate. For a simple example, consider a 2D antiferromagnet with a single propagation vector $\bq=(\pi,\pi)$ (Fig.\ref{fig:structure}(a)). This means $\mathbf{b}^M_1=(\pi/2,\pi/2)$ and $\mathbf{b}^M_2=(\pi/2,-\pi/2)$, and therefore A-TRIM are $(\pi/4,\pm\pi/4)$ and B-TRIM are $(0,0)$, which is always a B-TRIM, and $(\pi/2,0)$ corresponding to $(y_1,y_2)=(1,1)$.

\section{Kramers' Degeneracy in an AFM Insulator: Proof by Perturbation}\label{sec:Perturbation}

In the previous section, we have proved, by using only properties of symmetry operators, that a new type of Kramers' degeneracy is protected by a combined symmetry $\Theta_S$. The cases of single propagation vectors $\bq=(m\mathbf{b}_1+n\mathbf{b}_2+l\mathbf{b}_3)/2$, where $m,n,l$ has no common divisor, has been discussed in Ref.[\onlinecite{Mong2010}], but here we extended it to arbitrary commensurate AFM. The validity of the statement is independent of the particular strength of AFM or the specific magnetic structure. We here give a more intuitive approach to understanding this result. We start from another perspective by treating the AFM as a perturbation field applied to an otherwise TRS system, and study how the perturbation breaks/preserves the Kramers' degeneracy of the TRS system.

The single particle Hamiltonian of Eq.(\ref{eq:first_eq}) can be rewritten as
\bea
\hat{H}&=&\sum_{\bk,\alpha,\beta,\tau,\tau'}\mathcal{H}_{\alpha\tau,\beta\tau'}(\bk)c^\dag_{\alpha\tau}(\bk)c_{\beta\tau'}(\bk)\\
\nonumber&+&\sum_{\bk,i,\alpha,\tau,\tau'}(\mathbf{M}(\bQ_i)\cdot\sigma_{\tau\tau'})c^\dag_{\alpha\tau}(\bk+\bQ_i)c_{\alpha\tau'}(\bk).
\eea
where $\mathbf{M}(\bQ_i)$'s are the Fourier components of the Zeeman field $\mathbf{J}(\br)$. 
\begin{figure}
\includegraphics[width=8cm]{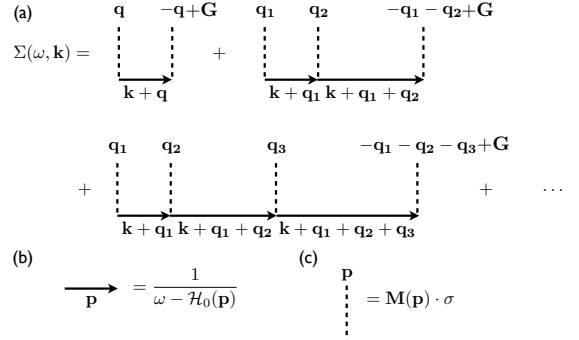}
\caption{(a) is the Feynman diagram expansion of the self-energy induced by the scattering of AFM field. (b) represents the free electron propagator in the momentum space without the AFM field. (c) is the vertex representing the scattering by the AFM field of momentum $\mathbf{p}$.}
\label{fig:feynman}
\end{figure}
The energy levels at $\bk$ present themselves as the poles of the single-particle Green's function:
\bea\label{eq:greendef}G(\omega,\bk)=\frac{1}{\omega-\mathcal{H}_0(\bk)-\Sigma(\omega,\bk)},\eea where $\Sigma(\omega,\bk)$ is the self-energy. When the AFM field is weak, the self-energy can be expanded in terms of the number of scatterings of an electron whose initial and final momentum are $\bk$ plus a reciprocal lattice vector $\mathbf{G}$. This scattering process can be visualized using Feynman diagrams, as shown in Fig.\ref{fig:feynman}(a), where the first three terms represent the electron being scattered by the AFM field once, twice and three times before returning to the initial state. Applying the conventional Feynman rules given in Fig.\ref{fig:feynman}(b,c), the self-energy is given by perturbation series:\begin{widetext}
\bea\label{eq:SigmaExpansion}
\Sigma(\omega,\bk)&=&\Sigma_1(\omega,\bk)+\Sigma_2(\omega,\bk)+...+\Sigma_n(\omega,\bk)+...\\
\nonumber\Sigma_n(\omega,\bk)&=&\sum_{\bq_i\in\{\bQ_1,...,\bQ_m\},\bq_1+..+\bq_n=\mathbf{G}}[\mathbf{M}(\bq_1)\cdot\sigma]G_0(\omega,\bk+\bq_1)...[\mathbf{M}(\bq_{n-1})\cdot\sigma]G_0(\omega,\bk+\bq_1+...+\bq_{n-1})[\mathbf{M}(\bq_n)\cdot\sigma]
\eea \end{widetext}where the free Green's function
\bea G_0(\omega,\bk)=\frac{1}{\omega-\mathcal{H}_0(\bk)}.\eea

We note that for the Green's function, $\omega$ should have an infinitesimal imaginary part, but since we are only interested in the position of \emph{real} poles, we take $\omega$ to be a real number. Due to the TRS of the unperturbed Hamiltonian, all poles of the free Green's function $G_0(\omega,\bk)$ are doubly degenerate. If there is a splitting of the degeneracy, it must be due to the self-energy term $\Sigma(\omega,\bk)$. Though in general $\Sigma(\omega,\bk)$ is not hermitian, it is hermitian in this problem as only elastic scattering is present (and using $\mathbf{M}(\bQ_i)=\mathbf{M}^\ast(-\bQ_i)$):\begin{widetext}
\bea
\nonumber\Sigma^\dag_n(\omega,\bk)&=&\sum_{\bq_i\in\{\bQ_1,...,\bQ_m\},\bq_1+..+\bq_n=\mathbf{G}}[\mathbf{M}(-\bq_n)\cdot\sigma]G_0(\omega,\bk+\bq_1+...+\bq_{n-1})...[\mathbf{M}(-\bq_2)\cdot\sigma]G_0(\omega,\bk+\bq_1)[\mathbf{M}(-\bq_1)\cdot\sigma]\\
\nonumber&=&\sum_{-\bq_i\in\{\bQ_1,...,\bQ_m\},-\bq_n-...-\bq_1=\mathbf{G}}[\mathbf{M}(-\bq_n)\cdot\sigma]G_0(\omega,\bk-\bq_n)...[\mathbf{M}(-\bq_2)\cdot\sigma]G_0(\omega,\bk-\bq_n-...-\bq_2)[\mathbf{M}(-\bq_1)\cdot\sigma]\\
&=&\Sigma_n(\omega,\bk).\eea\end{widetext} Therefore all poles of $G(\omega,\bk)$, like those of $G_0(\omega,\bk)$, are still real.

We will now focus on the symmetry property of each term in the expansion of $\Sigma(\omega,\bk)$. Under time-reversal operator of Eq.(\ref{eq:Theta1}), the free Green's function transforms as
\bea
\hat\Theta{G}_0(\omega,\bk)\hat\Theta^{-1}=G_0(\omega,-\bk).
\eea 
The AFM vertex from the expansion transforms as
\bea\hat\Theta\mathbf{M}(\bQ_i)\cdot\sigma\hat\Theta^{-1}=-\mathbf{M}(-\bQ_i)\cdot\sigma.\eea Therefore the $n$-th order self-energy transforms according to
\bea\label{eq:SigmaTransform}\hat\Theta\Sigma_n(\omega,\bk)\hat\Theta^{-1}=(-1)^n\Sigma_n(\omega,-\bk).\eea
Since for $n\in{even}$, $\hat\Theta\Sigma_n(\omega,\bk)\hat\Theta^{-1}=\Sigma_n(\omega,-\bk)$, then if $\Sigma_n=0$ for any $n\in{odd}$, we have
\bea\label{eq:SigmaTRS}\hat\Theta\Sigma(\omega,\bk)\hat\Theta^{-1}=\Sigma(\omega,-\bk).\eea Using Eq.(\ref{eq:SigmaTRS}) in conjunction with Eq.(\ref{eq:Theta1}), we know for each eigenstate $|u(\omega,\bk)\rangle$ of $\mathcal{H}_0(\bk)+\Sigma(\omega,\bk)$, $\hat\Theta|u(\omega,\bk)\rangle$ must be an orthogonal eigenstate of $\mathcal{H}_0(-\bk)+\Sigma(\omega,-\bk)$ with the same eigenvalue. Therefore, as far as $n\in{odd}$ terms are vanishing in the self-energy expansion, the AFM as a perturbation will \emph{not} split the Kramers' degeneracy in a TRS system. Physically, this can be easily understood as follows: if an electron must experience an \emph{even} number of scatterings by the AFM field before going back to its initial state, the scattering amplitude does not change under time-reversal which changes the sign of AFM; therefore the single electron propagator cannot `see' the breaking of TRS and its poles remain doubly degenerate as in the case where TRS is preserved.

We now determine the condition under which the odd terms in the self energy disappear. Let us start from an AFM of a simple type in a 1D insulator to gain some intuition. We assume that there is only one wavevector $p=2\pi/s$, where $s>1$ is an integer. In the $n$-th ($n\in{odd}$) order term of $\Sigma(\omega,\bk)$, each $q_i$ must take either $p$ or $-p$, but because of the constraint $\sum_iq_i=G$, we can easily see that if $s=even$, it is impossible to satisfy the constraint with an odd number of $q_i$'s, or in other words, $\Sigma_{n\in{odd}}(\omega,\bk)=0$. Moreover, if $s\in{odd}$, there are two cases: The first case is when $n<s$ where it is still not possible to satisfy the constraint if $n\in{odd}$, \ie $\Sigma_{n\in{odd},n<s}(\omega,\bk)=0$. The second case is when $n\ge{s}$, in this situation one may choose: $q_1=...=q_s=p$ and $q_{s+k}=(-1)^kp$ for $0<k\le n-s$ (since $n-s=even$ the sum over $q_{s+k}$ vanishes), to satisfy the constraint if $n\in{odd}$, \ie $\Sigma_{n\in{odd},n\ge{s}}(\omega,\bk)\neq0$, thereby breaking the degeneracy. Thus, by use of this simple example, we know that if $s=even$, all odd terms in $\Sigma(\omega,\bk)$ must vanish and if $s\in{odd}$, the lowest nonvanishing odd order is the $s$-th order.

A similar analysis can be made for the general case. In a general commensurate AFM, there are a finite number of propagation vectors denoted by $\bQ_1,...,\bQ_m$. At each scattering, there is $\bq_i\in\{\bQ_1,...,\bQ_m\}$, so a generic scattering process that contributes to the self energy has $z_i$ times of momentum transfer $\bQ_i$ and satisfies $\sum_iz_i\mathbf{Q}_i=\mathbf{G}$. Therefore, if we want all odd orders in $\Sigma(\omega,\bk)$ to vanish, we require that for any set of integers $z_{i=1,...,m}$ satisfying $\sum_iz_i\mathbf{Q}_i=\mathbf{G}$, $\sum_iz_i=even$. This is the sufficient and necessary condition under which every pair of doubly degenerate poles of $G_0(\omega,\bk)$ remains degenerate in the poles of $G(\omega,\bk)$, given that the AFM field is small enough to be treated as perturbation. (Unlike in the simple 1D example, however, when the condition is not satisfied, the general formula for the lowest order of scattering that breaks the degeneracy is unknown to us.)

In this and the previous section, we have established two \emph{equivalent} sufficient and necessary conditions for the preservation of Kramers' degeneracy in an AFM TI, which are:\\
 
(i) there exists a lattice vector $\mathbf{D}$, the translation by which flips all spins in the AFM and \\

(ii) for any $m$ integers satisfying $\sum_iz_i\bQ_i=\mathbf{G}$, then $\sum_iz_i=even$. \\

The actual proof of their equivalence is purely mathematical and is hence deferred to the Appendix. Here we comment that condition (i) is a direct extension of the case discussed in the earlier work by Mong \emph{et al}\cite{Mong2010}, where AFM field with propagation vector $(0,0,\pi)$, $(0,\pi,\pi)$ or $(\pi,\pi,\pi)$ satisfies this condition. Condition (ii) is a constraint on the propagation vectors and these vectors alone, making explicit the fact that the preservation of Kramers' degeneracy has nothing to do with the magnitudes or directions of $\mathbf{M}(\bQ_i)$.
 
Besides relating the propagation vectors to the degeneracy in the energy spectrum, the analysis of perturbation helps estimate the magnitude of the energy splitting if condition (i) or (ii) is violated. To see this, we return to the simplest example of 1D AFM with wavevector $p=2\pi/s$. We have mentioned that when $s\in{odd}$, the lowest order of self-energy that breaks the Kramers' degeneracy is the $s$-th order:\begin{widetext}
\bea\Sigma_s(\omega,k)=(\mathbf{M}(p)\cdot\sigma)G_0(\omega,k+p)(\mathbf{M}(p)\cdot\sigma)G_0(\omega,k+2p)...(\mathbf{M}(p)\cdot\sigma)G_0(\omega,k+(s-1)p)(\mathbf{M}(p)\cdot\sigma).\eea
\end{widetext}
The magnitude of this term is estimated as:
\bea
&&\Sigma_s(\omega,k)\\\nonumber&\sim&\frac{|M|^s}{(\omega-E(k+p))(\omega-E(k+2p))...(\omega-E(k+(s-1)p))}\\
\nonumber&\sim&\frac{|M|^s}{\prod_{i=1,...,s-1}|E(k)-E(k+ip)|}.
\eea
To go from the second to the last line, remember that since we are interested in the correction to the position of the real pole which is at $\omega=E(k)$ without AFM, we can approximately substitute $\omega=E(k)$ into the second line.

From this analysis, we can see that, assuming one can treat AFM exchange field perturbatively, the splitting becomes exponentially small as $s$, or the period, increases, justifying the intuition that when $s$ is large enough, there should be no difference between even $s$ (no splitting) and odd $s$ (exponentially small splitting).

One can always use the same analysis to estimate the splitting with a more complex magnetic structure. The key point is to find the first term in the series expansion that has odd number of $\mathbf{M}(\bq)$'s. If the condition (ii) does not apply, there exists among the sets of ${z_i}$ for which $\sum_i
z_i Q_i=\mathbf{G}$ and $\sum_i z_i\in{}odd$, a set for which $k\equiv\sum_i |z_i|$ is
minimal and the energy splitting is
\bea\label{eq:deltaE}
\Delta{E}\propto|M|^r.
\eea
For the simplest example, suppose there are two propagation vectors satisfying $Q_1=2Q_2=\pi/4$. Both vectors have even denominators, but since $2Q_1-Q_2=0$, three scatterings (odd number) send an electron to its original momentum, so the energy splitting should be $|M|^3$, not zero.

\begin{figure}[!htb]
\includegraphics[width=8cm]{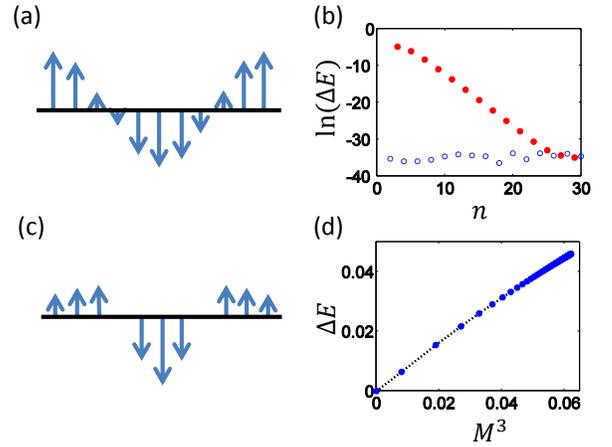}
\caption{(a)Magnetic structure of a sinusoidal antiferromagnet with wavevector $Q=\pi/5$ described in Eq.(\ref{eq:1DAFM}). (b) The energy splitting of the Kramers pair in a 1D TRS insulator in the presence of AFM shown in (a) at $k=0$ as a function of the spatial period $s$ of the AFM. Red dots correspond to odd and blue empty circles correspond to even $s$'s. (c) Magnetic structure given by Eq.(\ref{eq:1Dstructure}) with $s=10$. (d) The energy splitting of the Kramers pair in a 1D TRS insulator in the presence of AFM shown in (c) at $k=0$ plotted against the AFM strength cubed, $M^3$.}
\label{fig:1Dmodel}
\end{figure}

To solidify our understanding of the energy splitting, we use explicit models of 1D insulators with commensurate AFM. We take a basic four-band model for the TRS part of the 1D insulator:
\bea
\mathcal{H}_0(k)=(m-\cos(k))\Gamma_0+\sin(k)(\Gamma_1+b\Gamma_{15}),
\eea
where $\Gamma_0=1\otimes\tau_z$, $\Gamma_1=\sigma_x\otimes\tau_z$, $\Gamma_2=1\otimes\tau_y$, $\Gamma_3=\sigma_y\otimes\tau_x$, $\Gamma_5=\prod_{i=0,1,2,3}\Gamma_i$ and $\Gamma_{ij}=i\Gamma_i\Gamma_j$. The time-reversal operator is $\Theta=K\Gamma_{15}$. For the AFM part of the Hamiltonian, we consider the magnetic structure of a sinusoidal spin density wave with the period of $s$ lattice spacings (see Fig.\ref{fig:1Dmodel}(a)):
\bea\label{eq:1DAFM}
\mathbf{J}(x)=M\hat{z}\cos\frac{2\pi{x}}{s}.
\eea
Taking the parameters $m=0.5$, $b=0.01$ and $M=0.1$, we calculate the energy splitting of the Kramers' degeneracy at $k=0$ in the presence of the AFM field, and plot it as a function of $n$ in Fig.\ref{fig:1Dmodel}(b). From this Fig.\ref{fig:1Dmodel}(b), it is clear that when $s=even$, the energy splitting is zero (considering the numerical error) and when $s\in{odd}$, the energy separation decays exponentially with increasing $s$ for $s\gg1$, as predicted in Eq.(\ref{eq:deltaE}). Next we consider a magnetic structure with two harmonics, as shown in Fig.\ref{fig:1Dmodel}(c), given by
\bea\label{eq:1Dstructure}
\mathbf{J}(x)=M\hat{z}(\cos\frac{2\pi{x}}{s}+\frac{1}{2}\cos\frac{4\pi{x}}{s}).
\eea
Since $Q_1=2\pi/s$ and $Q_2=4\pi/s$, there is $Q_1-2Q_2=0$. This means that an electron at momentum $\bk$ can return to the initial state by three scatterings with the AFM field, contributing a TRS odd term to the self-energy which is proportional to $|M|^3$. Therefore, according to Eq.(\ref{eq:deltaE}), the energy splitting should be proportional to $|M|^3$. In Fig.\ref{fig:1Dmodel}(d), we plot the energy separation against $|M|^3$ for $s=10$ and we find that the result is consistent with the our analysis. 

As a final remark, all this perturbative analysis requires that $|M(\bQ_i)|$ be much smaller compared with the energy separation $|E_n(\bk)-E(\bk+\bQ_i)|$ and hence does not apply to flat band Hamiltonians where $E_n(\bk)=E(\bk+\bQ_i)$. The proof by symmetry presented in the previous section, however, applies to all cases.

\section{$Z_2$-Invariant for AFM Insulators}\label{sec:Z2}

In previous sections, we have derived the conditions under which the Kramers' degeneracy can be preserved in an AFM insulator. Specially, we have derived that at half of the TRIM (called B-TRIM) each energy level is doubly degenerate. In a TRS insulator the degeneracy at TRIM is key to defining the $Z_2$ topological invariant in 2D and 3D. Therefore, in this section, we seek to define the $Z_2$ invariant in an AFM insulator while restricting the discussion to dimensions $d\le3$.

We start by inspecting the sewing matrix associated with $\Theta_S$ symmetry at B-TRIM. Using Eq.(\ref{eq:ThetaS2}) and Eq.(\ref{eq:ThetaS3}), we have:
\bea\mathcal{B}_{mn}(\bk_{inv})&=&\langle\psi_m(\bk_{inv})|\ThetaS|\psi_n(\bk_{inv})\rangle\\
\nonumber&=&\langle\ThetaS^2\psi_n(\bk_{inv})|\ThetaS|\psi_m(\bk_{inv})\rangle\\
\nonumber&=&-\langle\psi_n(\bk_{inv})|\ThetaS|\psi_m(\bk_{inv})\rangle\\
\nonumber&=&-\mathcal{B}_{nm}(\bk_{inv}).\eea
This tells us that the sewing matrix is antisymmetric at all B-TRIM, which allows us to define the Pfaffian of the sewing matrix and, with it, define a $Z_2$ quantity $\delta_0$:
\bea\label{eq:Z21}(-1)^{\delta_0}=\prod_{\bk_{inv}\in\textrm{B-TRIM}}\frac{\Pf[\mathcal{B}(\bk_{inv})]}{\sqrt{\det[\mathcal{B}(\bk_{inv})]}}.\eea
For the $Z_2$ index in Eq.(\ref{eq:Z21}) to be well-defined, we need a smooth $\mathcal{B}(\bk)$ to have a consistent sign for the square root. This essentially requires that Stoke's theorem applies to any non-contractible closed loop in the MBZ, or equivalently, that any component of the Hall conductance $\sigma^H_{ij}$ must vanish. The Hall conductance can be calculated as\cite{Thouless1982}
\bea
\label{eq:Hall}
\nonumber\sigma_{ij}^H=\frac{i}{2\pi}\sum_{n\in{occ}}\epsilon_{ij}\int_{\textrm{2D plane in 3D BZ}}\langle\partial_iu_n(\bk)|\partial_ju_n(\bk)\rangle{}d^2k,\\
\eea 
where $|u_n(\bk)\rangle=e^{-i\bk\cdot{r}}|\psi_n(\bk)\rangle$ is the periodic part of the Bloch wave function. In an insulator, the Hall conductance, being quantized for any 2D plane in the 3D BZ, must be the same as we vary the out-of-plane component of $\bk$. Therefore, we can simply take this component of $\bk$ to be zero. Using Eq.(\ref{eq:ThetaS2}), we have\bea\epsilon_{ij}\langle\partial_iu_n(\bk)|\partial_ju_n(\bk)\rangle&=&\epsilon_{ij}\langle\partial_j\ThetaS{u}_n(\bk)|\partial_i\ThetaS{u}_n(\bk)\rangle\\
\nonumber&=&\epsilon_{ij}\langle\partial_ju_n(-\bk)|\partial_iu_n(-\bk)\rangle\\
\nonumber&=&-\epsilon_{ij}\langle\partial_iu_n(-\bk)|\partial_ju_n(-\bk).\eea
Substitute this equation into Eq.(\ref{eq:Hall}) taking the other $k_{m\neq{i,j}}=0$, we have
\bea
\label{eq:zeroHall}\sigma_{ij}^H=0.
\eea
Eq.(\ref{eq:zeroHall}) states that, for any combination of $i,j$, there is no obstruction to defining a smooth gauge for all occupied states in the MBZ, making Eq.(\ref{eq:Z21}) a meaningful definition.

However, this does not mean that $\delta_0$ defined in Eq.(\ref{eq:Z21}) is a topological invariant, because we have not proved that it is gauge invariant. To see this, we perform a gauge transform that is continuous in MBZ:
\bea|u'_n(\bk)\rangle=\sum_{m\in{occ}}{U}_{nm}(\bk)|u_m(\bk)\rangle,\eea where $U(\bk)$ is an arbitrary unitary matrix. The sewing matrix in the new basis becomes
\bea\label{eq:SewingTransform}\mathcal{B}'_{mn}(\bk)=\sum_{m'n'}U^\ast_{mm'}(-\bk)\mathcal{B}_{m'n'}(\bk)U^\dag_{n'n}(\bk).\eea From Eq.(\ref{eq:SewingTransform}) we have
\bea
\label{eq:temp6}\det[\mathcal{B}'(\bk)]&=&\det[U^\ast(-\bk)]\det[U^\dag(\bk)]\det[\mathcal{B}(\bk)]\\
\nonumber&=&(\det[U^\ast(\bk)])^2\det[\mathcal{B}(\bk)],\\
\nonumber\Pf[\mathcal{B}'(\bk_{inv})]&=&\det[U^\ast(\bk_{inv})]\Pf[\mathcal{B}(\bk_{inv})].\eea If in defining the complex square root function $\sqrt{z}$, we choose the branch cut such that if the argument of $z$ is within $[0,2\pi)$, we have that $\det[U^\ast(\bk_{inv})]/\sqrt{\det^2[U^\ast(\bk_{inv})]}=\pm1$ if $arg[\det[U^\ast(\bk_{inv})]]\in[0,\pi)$ and $arg[\det[U^\ast(\bk_{inv})]]\in[\pi,2\pi)$, respectively. If for odd number of $\bk_{inv}$ we have $arg[\det[U^\ast(\bk_{inv})]]\in[\pi,2\pi)$, the $Z_2$-number defined in Eq.(\ref{eq:Z21}) changes its value and is therefore gauge variant.

In a 1D $\Theta_S$-symmetric insulator, the only B-TRIM is $k=0$. If we can find a gauge choice with $arg[\det[U^\ast(0)]]\in[\pi,2\pi)$, the $Z_2$ number becomes gauge variant. Consider a constant gauge transform: $U_{mn}=\delta_{mn}(1-2\delta_{m1})$. By choosing this gauge we simply redefine the wave functions of the first band by multiplying a factor of $-1$ while the wave functions of the other bands remain the same. Therefore, we have $\det[U^\ast]=-1$. Under this new gauge the $Z_2$-number changes its value and is a gauge variant quantity, and hence no topological $\Theta_S$-symmetric insulators exist in 1D.

If one considers a 2D $\Theta_S$-symmetric insulator, there are three possible lattice translation vectors ($\mathbf{D}$'s): $\mathbf{D}_1=\mathbf{a}^M_1/2,\mathbf{D}_2=\mathbf{a}^M_2/2,\mathbf{D}_3=(\mathbf{a}^M_1+\mathbf{a}_2^M)/2$, where $\mathbf{a}_{1,2}^M$ are the basis vectors of the magnetic superlattice. The question comes down to if we can find a gauge satisfying $arg[\det[U^\ast(\bk_{inv})]]\in[\pi,2\pi)$ at one of the two B-TRIM. When $\mathbf{D}=\mathbf{D}_1$, there are two B-TRIM located at $\mathbf{K}_1=0$ and $\mathbf{K}_2=\mathbf{b}^M_2/2$. Consider the gauge transform
\bea
U_{mn}(\bk)=\delta_{mn}(1+(\exp(i\bk\cdot\mathbf{a}^M_2)-1)\delta_{m1}).
\eea 
In this gauge, the wave function of the first band is multiplied by a factor of $\exp(i\bk\cdot\mathbf{a}^M_2)$ while the wave functions of the other bands remain unchanged\cite{fu2007a}. Using Eq.(\ref{eq:temp6}) we easily obtain $\det[\mathcal{B}'(\mathbf{K}_1)]=1$, $\det[\mathcal{B}'(\mathbf{K}_2)]=-1$. This gauge choice changes the value of the $Z_2$-number. For the other two possible $\mathbf{D}$ vectors, one can choose the gauge $U_{mn}(\bk)=\delta_{mn}(1+(\exp(i\bk\cdot\mathbf{a}^M_1)-1)\delta_{m1})$ and $U_{mn}(\bk)=\delta_{mn}(1+(\exp(i\bk\cdot(\mathbf{a}^M_1-\mathbf{a}^M_2))-1)\delta_{m1})$ and reach the same conclusion. Therefore the $Z_2$-number as defined in Eq.(\ref{eq:Z21}) is gauge variant also in 2D.

The gauge variance in 1D and 2D can be physically understood by applying the established theory classifying TRS insulators. To see this, one notices that the equation $\bk\cdot\mathbf{D}=0$ defines a $d-1$-dimensional subsystem of the $d$-dimensional AFM insulator, in which we have
\bea[\ThetaS,\hat{H}_{d-1}]&=&0,\\
\nonumber\ThetaS^2=-\exp(2\bk\cdot\mathbf{D})&=&-1.\eea These two equations are the defining properties of the symplectic class\cite{Altland1997} (class AII). Hence, the subsystem equals a spinful TRS insulator in $d-1$-dimension. Since the topological classification of 0D and 1D symplectic insulators is trivial, the 1D and 2D $\Theta_S$-symmetric insulators must also be trivial. Therefore the $Z_2$-number as defined in Eq.(\ref{eq:Z21}) must be gauge variant in 1D and 2D, and no 1D and 2D $\Theta_S$-symmetric TI's exist.

In the same light, the gauge \emph{invariance} of the $Z_2$-number in 3D $\Theta_S$-symmetric AFM insulators is easy to understand. The Hamiltonian constrained to the 2D plane satisfying $\bk_{//}\cdot\mathbf{D}=0$ is exactly the same as a 2D TRS insulator with time-reversal symmetry replaced by $\Theta_S$, the combined symmetry, for which the $Z_2$-invariant given by Fu and Kane can be written down
\bea\label{eq:Z22}(-1)^{\delta_0}=\prod_{\bk_{inv}^{//}}\frac{\Pf[\mathcal{B}(\bk_{inv}^{//})]}{\sqrt{\det[\mathcal{B}(\bk_{inv}^{//})]}}.\eea There are eight TRIM and four B-TRIM in any 3D insulator, and due to $\exp(2i\bk^{//}_{inv}\cdot\mathbf{D})=1$, all four $\bk^{//}_{inv}$'s are B-TRIM in the 3D MBZ. As a result, the $Z_2$ quantity defined in Eq.(\ref{eq:Z22}) is exactly the same as the one defined in Eq.(\ref{eq:Z21}) for 3D AFM insulators. Since in 2D TRS insulators the $Z_2$ quantity defined in Eq.(\ref{eq:Z22}) is gauge-invariant, we conclude that the $Z_2$ quantity defined in Eq.(\ref{eq:Z21}) is also gauge invariant.

While we have proved that all $\Theta_S$-symmetric AFM insulators have Kramers' degeneracy at B-TRIM, only in 3D can we define the $Z_2$ topological invariant. However, when there are additional symmetries beyond $\Theta_S$ in the system, these invariants may be defined in lower dimensions\cite{hughes2010inv,Turner:2012,Fang2012}. For example, when spatial inversion symmetry is present in the system, we can have well-defined $Z_2$ invariants in all dimensions. To see this, assume $\hat{P}$ to be the inversion symmetry operator in the single-particle Hilbert space. The commutation relation between the inversion operator and an arbitrary translation operator is
\bea\hat{T}_\br\hat{P}=\hat{P}\hat{T}_{-\br},\eea
and the commutation relation between the inversion operator and time-reversal operator is
\bea\hat{\Theta}\hat{P}=\hat{P}\hat{\Theta}.\eea Therefore we have the commutation relation between $\ThetaS$ and $\hat{P}$ as
\bea\label{eq:PandThetaS}\hat{P}\ThetaS&=&\hat{P}\hat\Theta\hat{T}_\mathbf{D}\\
\nonumber&=&\hat\Theta\hat{T}_{-\mathbf{D}}\hat{P}\\
\nonumber&=&\ThetaS\hat{P}\hat{T}_{2\mathbf{D}}.\eea At B-TRIM, we have $\hat{T}_{2\mathbf{D}}|\psi_n(\bk_{inv})\rangle=|\psi_n(\bk_{inv})\rangle$. This means in the subspace spanned by the two degenerate states at any B-TRIM, the inversion operator and $\ThetaS$ commute. If one state has parity (inversion eigenvalue) $+1$ ($-1$), the other must have the same parity. This allows us to define the following $Z_2$ invariant $\zeta_0$:
\bea
(-1)^{\zeta_0}=\prod_{\bk_{inv}\in\textrm{B-TRIM},n\in{occ}/2}\zeta_n(\bk_{inv}),\eea
where $\zeta_n$ is the parity of the $n$-th occupied band and the multiplication over ${occ}/2$ means only one state in a Kramers' pair is chosen. This definition applies to 1,2 and 3D and can be extended using the same definition to any dimensions. These insulators, being stabilized by the added inversion symmetry, will, however, not exhibit gapless boundary modes in the energy spectrum but only in the entanglement spectrum.

\section{Gapless Boundary Modes}\label{sec:Boundary}

In 2D and 3D TRS insulators, there are odd number of Dirac boundary modes in $Z_2$ non-trivial insulators and even number (including zero) of Dirac boundary modes in $Z_2$ trivial insulators, i.e., on the surface Brillouin zone (SBZ) of a 3D TRS TI, there must be at least one Dirac node at one of the TRIM. At that Dirac point a Kramers' pair protected by time-reversal symmetry is located. Breaking TRS, in general, can gap the Dirac point. For example, when the Fermi energy is close to the Dirac point energy, a magnetic field in a 2D TRS TI localizes the edge modes\cite{Delplace2012}; and in a 3D TRS TI, applying an FM Zeeman field on the surface opens gap at each Dirac point and makes the surface a 2D quantum anomalous Hall insulator, with each gapped Dirac node contributing $\pm e^2/h$\cite{Chang2013}.

When an AFM field is applied to a TRS TI, however, the Dirac point at a B-TRIM is still doubly degenerate if the system, with the open surfaces, still possesses $\Theta_S$-symmetry. This requirement not only implies that the bulk preserves $\Theta_S$-symmetry, but also implies that the surface termination does not violate that symmetry. This essentially puts constraints on how the system should be terminated in real space.

We want to find out what types of surface terminations can preserve the Dirac nodes, \ie the $\Theta_S$-symmetry. First we remember that although we have been using $\mathbf{D}$ as if it were a unique vector, in fact $\mathbf{D}$ is unique only up to a magnetic lattice translation, i.e., any $\mathbf{D}'=\mathbf{D}+\mathbf{L}^M$ is also a translation vector that inverts all spins. We so far have made no distinction between them because in a periodic system, if $\hat\Theta_S=\hat\Theta\hat{T}_{\mathbf{D}}$ is a symmetry then $\hat\Theta'_S=\hat\Theta\hat{T}_{\mathbf{D}'}$ is also a symmetry and vice versa. For an open system, in order to preserve $\Theta_S$, we need at least one ${\mathbf{D}}$ that (i) inverts all spins and (ii) is parallel to the surface. Rewrite Eq.(\ref{eq:generalD}) after expanding $\mathbf{L}^M=\sum_in_i\mathbf{a}^M_i$
\bea\label{eq:Dform}
{\mathbf{D}}=\sum_{i=1,...,d}(n_i+\frac{x_i}{2})\mathbf{a}^M_i,
\eea 
where $n_i$ is an arbitrary integer and $x_i$ is either zero or unity with at least one $x_i$ being unity. (Here we assume that $d$ is a dimension in which there is non-trivial topological class for TRS insulators\cite{Kitaev2009}.)  If we use $\mathbf{N}$ to denote the normal direction of an open boundary, the protection of surface states require the existence of a set of $\{n_1,...,n_d\}$ in Eq.(\ref{eq:Dform}) such that
\bea\label{eq:condition}
\mathbf{N}\cdot{\mathbf{D}}=0.
\eea 

The above result may be understood from a more physical point of view. When the condition Eq.(\ref{eq:condition}) is met, all spins on the boundary change sign under an in-plane translation of ${\mathbf{D}}$, and hence the total magnetization on the boundary is zero. To the lowest order perturbation, the total magnetization, i.e., ferromagnetism, on the boundary opens a Zeeman gap at a TRIM. No gap is opened to first order when the total magnetization vanishes. On the other hand, if $\mathbf{N}$ is \emph{not} perpendicular to any $\mathbf{D}$, the total magnetization on the surface becomes finite and a gap is opened. The gap can be estimated by calculating the total magnetization within the decay length of the surface states. To be more concrete, let us consider a 2D AFM insulator, the TRS part of which is given by
\bea\label{eq:2Dmodel}
\mathcal{H}_0(k_x,k_y)&=&(m-\cos(k_x)-\cos(k_y))\Gamma_0\\
\nonumber&+&\sum_{i=x,y}\sin(k_i)(\Gamma_i+b\Gamma_{i5}).
\eea
\begin{figure}[!htb]
\includegraphics[width=8cm]{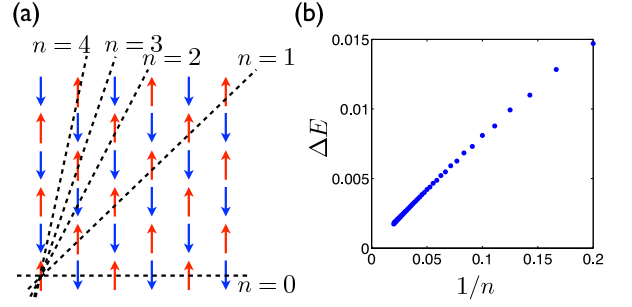}
\caption{(a) Different terminations of a 2D AFM insulator with propagation vector $(\pi,\pi)$, for which the coordinates of sites on the boundary satisfy $y=nx$. (b) The energy separation at the Dirac point of the system described by Eq.(\ref{eq:2Dmodel}) with $k_z=0$, $m=1.5$ and $M=0.2$, as a function of the inverse slope of the cut for $n\in{odd}$. If $n\in{even}$, the energy separation is zero.}
\label{fig:2Dmodel}
\end{figure}
This tight-binding model can be considered as representing the 2D TRS TI of HgTe for momenta near $\Gamma$ as $1<|m|<2$\cite{Bernevig:2006kx}. The configuration of the AFM field is given by (see Fig.\ref{fig:2Dmodel}(a))
\bea\label{eq:2Dstructure}
\mathbf{J}(x,y)=M\hat{z}\cos(\pi{x})\cos(\pi{y}).
\eea
It is easy to see that this AFM insulator has $\Theta_S$ symmetry, with $\mathbf{D}=(\mathbf{a}^M_1+\mathbf{a}^M_2)/2+\mathbf{L}^M$, where $\mathbf{a}^M_{1,2}=\mathbf{a}_2\pm\mathbf{a}_1$. We choose several boundaries (edges) with the form: \bea\mathbf{N}_n=-n\mathbf{a}_1+\mathbf{a}_2.\eea For $n\in{even}$, Eq.(\ref{eq:condition}) is satisfied, by finding ${\mathbf{D}}=n/2(\mathbf{a}^M_1-\mathbf{a}^M_2)+(\mathbf{a}^M_1+\mathbf{a}^M_2)/2=\mathbf{a}_1+n\mathbf{a}_2$; but if $n\in{odd}$, Eq.(\ref{eq:condition}) is unsatisfied and a gap is opened at the Dirac point. There are two ways to see the latter point: first, one can try to solve Eq.(\ref{eq:condition}) substituting $\mathbf{N}_n$ with $x_1=x_2=1$ and find that no integer solution for $n_i$'s exists, or one simply notices that along the cut, the spins are all ferromagnetically aligned, so it is impossible to find any translation along the edge that inverts all spins. Along the edge, the distance between two nearest sites is $\sqrt{n^2+1}a$, so the number of sites per unit length is $\rho=1/\sqrt{n^2+1}a\sim1/na$ and the distance between the first and the second layer of atoms is $d_n=a/\sqrt{1+n^2}\sim{a}/n$. The decay length of the edge mode at TRIM can be estimated as $l\sim{v}_F/\Delta_{bulk}$, where $v_F$ is the Fermi velocity at the Dirac point without AFM and $\Delta_{bulk}$ is the bulk insulating gap. The total magnetization close to the surface, or the ferromagnetic gap, is therefore
\bea\label{eq:estimate}
\Delta{E}_n&\propto&\rho(1-\exp(-d/l)+\exp(-2d/l)-...)\\
\nonumber&=&\rho/(1+\exp(-d/l))\\
\nonumber&\propto&{1/n},\eea when $n$ is large. By choosing the parameters $m=1.5$, $b=0.01$ and $M=0.2$, we calculate the energy separation of the Dirac point at $\bk=\bar\Gamma$ in the edge Brillouin zone (EBZ). In Fig.\ref{fig:2Dmodel}(b), the energy separation is plotted versus $1/n$, and the linearity of the curve confirms Eq.(\ref{eq:estimate}). This corroborates our intuition that when $n$ is large, a surface with $n\in{even}$ and $n\in{odd}$ cannot be physically distinguished.

We have so far discussed whether the boundary Dirac modes are stable against a perturbative AFM field. Does this hold beyond the regime of perturbation? Topological protection of gapless boundary modes in TI's means that: (1) the Dirac point is not split and (2) the boundary bands always lie in the bulk gap connecting the conduction and the valence bands through spectral flow. The second point is important because otherwise one can adiabatically push whole boundary band into the conduction or the valence bands, making the system undistinguishable from a trivial insulator. We will from now on focus on whether the boundary gapless modes can remain within the bulk gap, or if they can be adiabatically pushed into the conduction or the valence bands.

We begin by considering a 2D TRS TI subject to a perturbative AFM field that is $\Theta_S$-symmetric, in order to show from another perspective that AFM TI does not exist in 2D. We terminate the system and make an edge, the normal direction of which satisfies Eq.(\ref{eq:condition}) for some ${\mathbf{D}}$. Along the edge, a translation by ${\mathbf{D}}$ flips all spins, therefore we know the magnetic unit cell on the edge is of length $2{\mathbf{D}}$ and the edge Brillouin zone (EBZ) extends from $\frac{-\pi}{2{D}}$ to $\frac{\pi}{2{D}}$. Within the EBZ there are two TRIM: $K_1=0$ and $K_2=\frac{\pi}{2{D}}$. By the definition of A-TRIM and B-TRIM (see below Eq.(\ref{eq:ATRIM})), we know that only $K_1$ is a B-TRIM. Therefore, without additional symmetries, in the EBZ, the edge bands must be doubly degenerate at $K_1$ (zone center) and non-degenerate at $K_2$ (zone corner). The non-degeneracy at zone corner is the key difference of this edge band dispersion from the dispersion of edge bands of all 2D TRS TI's and prevents the existence of spectral flow from the conduction band to the valence band.

\begin{figure}[!htb]
\includegraphics[width=8cm]{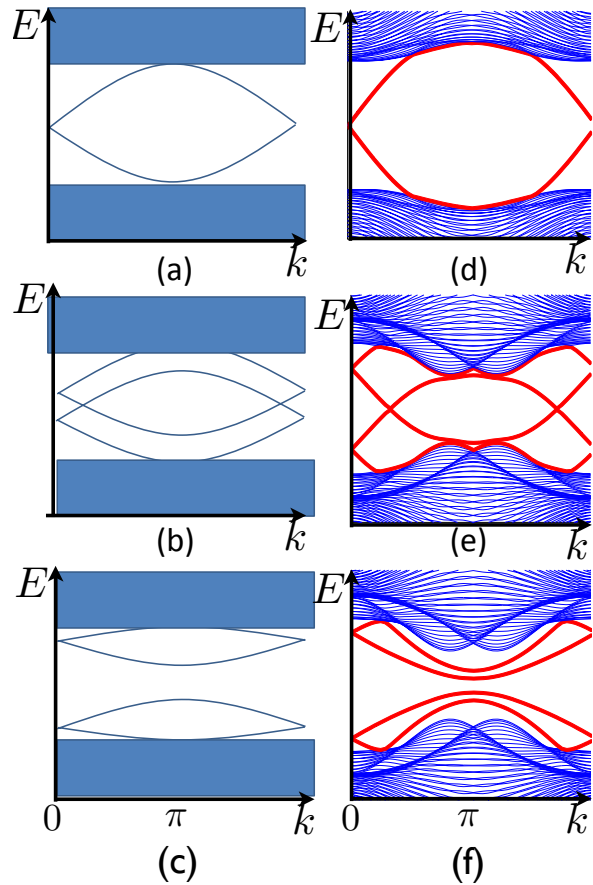}
\caption{(a-c) Edge state dispersion under different parameters without breaking the symmetries. It shows an adiabatic process, from (a) to (c), in which a full gap is open between the edge bands. (d-f) Edge and bulk dispersion of a 2D TRS TI described by Eq.(\ref{eq:2Dmodel}) subject to AFM given by Eq.(\ref{eq:2Dstructure}), in which the edge is along $y$-axis. From (d) to (f), the bias potential is increased to separate the edge bands without either breaking the symmetry or closing the band gap. In (d), $M=0.2$ (perturbative) and in (e,f), $M=0.4$ (non-perturbative).}
\label{fig:disentangle}
\end{figure}

In Fig.\ref{fig:disentangle}(a), a schematic edge dispersion is plotted. There are two Dirac cones because the dispersions of both edges (left and right) are shown. The topology of a band structure remains unchanged as one arbitrarily distorts its shape without breaking any protected degeneracy. In our case, without a degeneracy at the zone corner, one can distort the dispersion from Fig.\ref{fig:disentangle}(a) to Fig.\ref{fig:disentangle}(b) and finally to Fig.\ref{fig:disentangle}(c) where a full gap is obtained, without closing the bulk gap. Since we have always preserved the degeneracies at the zone center, the fully gapped system in Fig.\ref{fig:disentangle}(c) is topologically undistinguishable from the one in Fig.\ref{fig:disentangle}(a) where we have gapless edge modes. From this simple picture, we conclude that in a 2D AFM insulator, there is no protected gapless edge mode, although in the perturbation regime, the Dirac points of a TRS TI do not split and remain gapless.

We explicitly realize the scenario sketched in Fig.\ref{fig:disentangle}(a-c) in the 2D model given in Eq.(\ref{eq:2Dmodel}), but with a magnetic structure
\bea\label{eq:2Dstructure}
\mathbf{J}(x,y)=M\hat{z}\cos(\pi{y}).
\eea
The system is cut along $y$-direction and according to Eq.(\ref{eq:condition}), the degeneracy at $\Gamma$ in the EBZ is preserved. If we choose $M=0.2$, the dispersion, plotted in Fig.\ref{fig:disentangle}(d), is similar to that of the edge mode of a TRS 2D TI. Then we increase the strength of the AFM field to $M=0.4$ (see Fig.\ref{fig:disentangle}(e)) and add different chemical potentials on the left and the right edges $\mu_L=-\mu_R=0.5$, the dispersion becomes the one shown in Fig.\ref{fig:disentangle}(f), exhibiting a full gap between the conduction and the valence bands.

In 3D, consider a system with a surface having normal vector $\mathbf{N}$. We can define the magnetic supercell, spanned by $\mathbf{a}^M_{1,2,3}$ such that $\mathbf{a}^M_{1,2}$ are parallel to the surface, i.e., perpendicular to $\mathbf{N}$. (Generally speaking, such a choice breaks point group symmetries of the original lattice, but these symmetries are irrelevant in this paper.) Defining $\mathbf{b}^M_{1,2,3}$ from $\mathbf{a}^M_{1,2,3}$, we see that the SBZ is spanned by $\mathbf{b}^M_{1,2}$, with three TRIM being $\bar\Gamma=0$, $\bar{X}=\mathbf{b}^M_1/2$, $\bar{Y}=\mathbf{b}^M_2/2$ and $\bar{M}=(\mathbf{b}^M_1+\mathbf{b}^M_2)/2$.

Since ${\mathbf{D}}$ is parallel to the surface, it must be equal, up to an in-plane superlattice vector, to one of the following translation vectors: $\mathbf{D}_1=\mathbf{a}^M_1/2$, $\mathbf{D}_2=\mathbf{a}^M_2/2$ or $\mathbf{D}_3=(\mathbf{a}^M_1+\mathbf{a}^M_2)/2$. According to the definitions of A-TRIM and B-TRIM we find that the B-TRIM are $\bar\Gamma$, $\bar{Y}$ when ${\mathbf{D}}=\mathbf{D}_1$. Therefore, if we look at the band structure along $\Gamma{Y}$, there will be a protected surface mode always inside the bulk gap. On the other hand, when we look at the band structure along $\bar\Gamma\bar{X}$, we recover the situation discussed for 2D AFM insulators, in which the surface band is doubly degenerate at $\bar\Gamma$ but non-degenerate at $\bar{X}$, and can therefore be fully gapped by deformation. However, since along $\bar\Gamma\bar{Y}$ the surface mode always connects the conduction and the valence bands, the surface band as a whole is still topologically protected, and can never be pushed into the conduction or the valence band. The B-TRIM are $\bar\Gamma$, $\bar{X}$ when ${\mathbf{D}}=\mathbf{D}_2$, and $\bar\Gamma$, $\bar{M}$ when ${\mathbf{D}}=\mathbf{D}_3$. Similar statements on the surface modes can be made for these two cases by knowing the B-TRIM. To summarize the discussion of 3D AFM TI, we have proved that for an open surface that does not break the $\Theta_S$-symmetry, there is an odd number of surface bands crossing the band gap. The dispersion of these surface bands differ from those of 3D TRS TI in that along the direction having $\exp(2i\bk\cdot\mathbf{D})=1$, the dispersion runs from the valence to the conduction band just like the surface band of a TRS TI, but along the other direction, the dispersion does not touch the conduction or the valence band.

It should be noted that the statement on the existence of protected boundary modes runs completely in parallel to the statement concerning whether a $Z_2$-invariant can be defined. In a 2D AFM insulator, one cannot define a $Z_2$-invariant because the quantity defined in Eq.(\ref{eq:Z21}) is gauge variant in 2D, while in this section we have proved that it does not have protected edge modes that exhibit spectral flow, unlike 2D TRS TI, either. In a 3D AFM insulator, one can define a $Z_2$-invariant by Eq.(\ref{eq:Z21}) and at the same time, when it is $Z_2$-non-trivial, there is an odd number of protected surface modes inside the bulk gap.

\begin{figure}[!htb]
\includegraphics[width=8cm]{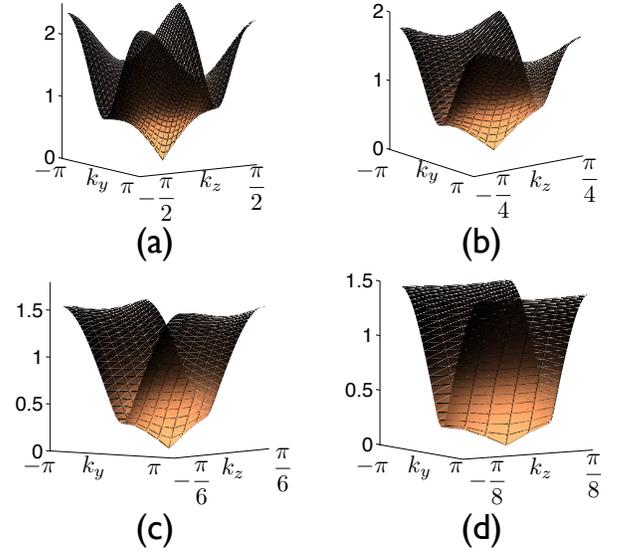}
\caption{The dispersion of the central surface band in the MBZ on surface in a 3D AFM TI with magnetic structure given by Eq.(\ref{eq:3Dstructure}) with $n=2,4,6,8$, respectively. The spectral weight for the first two layers of the surface is overplotted with the dispersion.}
\label{fig:dispersion}
\end{figure}

Below we use an explicit model to demonstrate the above statements. The TRS part of the 3D model is given by
\bea
\label{eq:3D}\mathcal{H}_0(\bk)&=&(m-\sum_i\cos{k_i})\Gamma_0\\
\nonumber&+&\sum_{i}\sin{k_i}(\Gamma_i+b\Gamma_{i5}),
\eea
with parameters $m=2.5$ and $b=0.01$. This tight-binding model is equivalent to the one for Bi$_2$Se/Te$_3$ in the continuum limit\cite{zhang2009}, and on each open surface has a Dirac cone centered at $\Gamma$ of the SBZ. The AFM field is given by
\bea\label{eq:3Dstructure}
\mathbf{J}(z)=M\hat{z}\cos\frac{2\pi{z}}{n}.
\eea
We choose the $yz$-plane as the open surface. Based on Eq.(\ref{eq:condition}), we know that the Kramers' pair at $\bar\Gamma$ of the SBZ is preserved if $n\in{even}$. The dispersion on the SBZ for $n=2,4,6,8$ are plotted in Fig.\ref{fig:dispersion}(a-d) for $M=0.2$. From these figures, we can see that the band dispersion along $y$-axis and that along $z$-axis are very different because $\bar{Y}$ is a B-TRIM and $\bar{Z}$ an A-TRIM. Along $y$-axis, the band dispersion always cross from the valence band to the conduction band, while along $z$-axis, the dispersion reaches a maximum energy below the conduction band; and the maximum energy decreases as $n$ increases.
A more experimentally accessible quantity than the energy dispersion is the electron spectral weight that can be directly measured in an angle resolved photoemission spectroscopy\cite{Hsieh2009,hsieh2009a,Xu2012} (ARPES),
\bea\label{eq:specweight}
\nonumber A(\omega,\bk)=\sum_{\alpha,\tau,n}\im\int{dt}\langle{c}_{n,\alpha\tau}(\bk)c^\dag_{n,\alpha\tau}(\bk,t)\rangle\exp(i\omega{t}),\\
\eea where $n$ denotes the layer the electronic operator is acting on (because the translational symmetry has been broken in the stacking direction). In reality, ARPES is a surface probe and therefore cannot penetrate an arbitrary number of layers of atoms, and here we assume that it only probes the spectral weight of the \emph{surface} rather than the bulk states. For a rough estimate, we only add up the contribution from the first two layers in Eq. (\ref{eq:specweight}). In Fig.\ref{fig:dispersion}, the surface spectral weight is over-plotted with the band dispersion, where lighter/darker means higher/lower spectral weight. From these figures, we can see that for AFM TI with large wavevector (small $n$), e.g., $n=2$, the spectral weight seen in experiment differs only minutely from what can be seen on the surface of a 3D TI, because the zone corner where the band bends backwards is actually filled with states distributed mainly in the bulk. As $n$-increases, the surface states begin to show some downward bending along the $z$-axis. 
\begin{figure}[!htb]
\includegraphics[width=8cm]{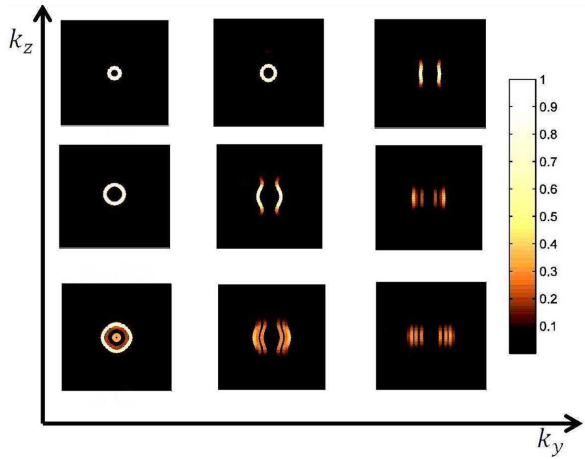}
\caption{The $\bk$-resolved spectral weight on the open surface of a 3D AFM TI with a magnetic structure given in Eq.(\ref{eq:3Dstructure}). The spectral weight is plotted for $E=0.3,0.5,0.7$ in Row 1,2 and 3 respectively. The first column corresponds to $M=0$, that is, a TRS TI; the second one corresponds to $M=0.2$ and $n=4$; the third column corresponds to $M=0.2$ and $n=8$. The color bar shows the distribution of a state on the first two layers of the surface where unity means completely bound to the surface.}
\label{fig:arpes}
\end{figure}

In Fig.\ref{fig:arpes}, we have plotted the spectral weight at certain energies, assuming that ARPES can only pick up spectral weight from the first two layers. From these figures we can see that if there is no AFM, the equal energy contour is always closed, but when $M\neq0$, at very low energy it is still closed but becomes open in the $z$-direction as one increases the energy. The critical energy at which the FS becomes open decreases as $n$ increases (or the wavevector decreases). As opposed to the equal energy contours in the dispersion plotted in Fig.\ref{fig:dispersion} which are always closed, the intensity plot of the spectral weight can show open arcs. This is because the states far away from the Dirac point are mostly bulk states having small spectral weight on the surface.

\section{Discussion}
\label{sec:discussion}
What are the possible realizations of an AFM TIs? In Ref.[\onlinecite{Mong2010}], GdBiPi was suggested as a TI with intrinsic AFM ordering due to the presence possibly strong and frustrating spin exchange coupling. Experimentally, there has been no report of any TI that is AFM ordered. On the other hand, in principle, we may use AFM substrate to induce AFM ordering in a TI thin film. For example, substrates of under-doped cuprates and the parent compounds of iron-based superconductors have strong AFM orderings and can induce AFM in the TI thin film with propagation vector $(\pi,\pi,0)$ and $(\pi,0,0)$, respectively\cite{Tsuei2000,Cruz2008}. For more complex AFM patterns one can use substrates with spiral antiferromagnetism or sinusoidal antiferromagnetism. In practice, we need the lattice matching between the substrate and the TI. For example, AFM substrates with tetragonal (cuprates) and orthogonal (iron-based superconductors) lattices in general do not match the rhombohedral lattice of Bi$_2$Se/Te$_3$. But based on the fact that the lattice constants in the 122-family of iron-based superconductors ($\sim4\AA$) are close to the lattice constant of the recently suggested TI of SmB$_6$\cite{Lu2013,Wolgast2012,Botimer2012} ($\sim4.13\AA$), we speculate that one may use the AFM state of 122 iron-based superconductors to induce stripe-like AFM in a thin film of SmB$_6$. If the lattices do not match, we need them to be at least commensurate with each other. In that case, at the interface, one needs to redefine superlattice, such that under translations by the redefined superlattice vectors, both lattices are invariant. Then we can calculate the propagation vectors in the new basis and check if $\Theta_S$-symmetry is preserved. If it is, from the theory we know that the AFM preserves the Dirac degeneracy of the surface states, but makes the dispersion anisotropic as shown in Fig.\ref{fig:arpes}.

Part of the discussion for AFM insulators can be easily extended to superconductors with coexisting AFM orderings. Specially, suppose we consider the superconductors with TRS and spin-orbital coupling (class DIII), and with an intrinsic or induced AFM coexisting with SC. If the system is $\Theta_S$-symmetric, the $d-1$-dimensional subsystem defined by $\bk\cdot\mathbf{D}=0$ belongs to class DIII, where the TRS is replaced by $\Theta_S$ which for this subsystem satisfies $\hat\Theta^2_S=-1$. Therefore a $d$-dimensional AFM SC with $\Theta_S$ symmetry has the same topological classification as a $d-1$-dimensional TRS superconductors (class DIII). From the general classification in Ref.\onlinecite{Schnyder2008}, 2D and 1D DIII superconductors have $Z_2$ classification; therefore, 3D and 2D $\Theta_S$-symmetric superconductors also have $Z_2$ classification. The state with coexisting AFM and SC may be realized by inducing an AFM in a 3D TSC thin film; for 3D TSC materials, copper doped Bi$_2$Se/Te$_3$\cite{Fu2010} and indium doped SnTe\cite{Sasaki2012} have been proposed. We have assumed that no symmetry exists other than $\Theta_S$, but in general AFM materials can have non-trivial magnetic groups, which possibly lead to much richer topological classifications beyond the scope of the current work.

\section{Conclusion}\label{sec:Conclusion}
In this paper, we carry out a general study of the role of commensurate antiferromagnetism in topological insulators, generalizing the previous results of Ref.\onlinecite{Mong2010}. The Hamiltonian of the electrons is modeled by a general TRS tight-binding model subject to a space dependent periodic Zeeman field having zero average (zero ferromagnetism). It is found that when there exists a lattice translation $\mathbf{D}$ that reverses the AFM field at every site, the Kramers' pairs at half of the TRIM are unbroken by the AFM field, while at the other half of the TRIM the eigenvalues are generically non-degenerate. The existence of $\mathbf{D}$ is guaranteed when the Fourier modes of the AFM field satisfies the following condition: for any set of $m$ integers $z_{1,...,m}$, if $\sum_{i=1,...,m}z_i\bQ_i=\mathbf{G}$, then $\sum_{i=1,...,m}z_i$ must be even, where $\bQ_i$'s are the propagation vectors of the AFM field. If this condition is not met, the energy splitting of the Kramers' pair can be estimated by knowing the propagation wavevectors of the AFM field. Furthermore, we show that the Kramers' pairs at half of the TRIM make possible the definition of a new $Z_2$-quantity similar to the one defined for TRS insulators. However, the new $Z_2$-index is only meaningful in 3D, but not in 1D or 2D (as it becomes a gauge variant quantity). The surface manifestation of the new $Z_2$-index is the presence of gapless Dirac modes in 3D AFM TIs. There are three differences between the surface modes of a AFM TI and those of a TRS TI: (i) to see the surface states, in AFM TI the surface of termination must preserve the combined symmetry $\Theta_S$ while in TRS TI the system can be terminated at any surface; (ii) in a AFM TI, the dispersion of the surface band only connects the conduction and valence bands in one direction, but is separated from the bulk bands along the other direction, while in a TRS TI the surface band connects the conduction and the valence bands along any direction; (iii) the spectral weight of the surface states at a certain in-gap energy makes a closed loop in a TRS TI, but appears as disconnected arcs in an AFM TI at energies away from the Dirac point. We discuss the materials in which the theory can be applied and also how the theory can be easily extended to the case of topological superconductors with coexisting AFM orders. Since in our work we have not assumed any other symmetry or any specific magnetic structure, the theory applies to general insulators with commensurate AFM.

\begin{acknowledgments}
CF is supported by ONR - N0014-11-1-0728, and acknowledges travel support from ONR - N00014-11-1-0635 and David and Lucile Packard Foundation. MJG acknowledges support from the AFOSR under grant FA9550-10-1-0459 and the ONR under grant N0014-11-1-0728. BAB is supported by NSF CAREER DMR- 095242, ONR - N00014-11-1-0635, Darpa - N66001-11- 1-4110, and David and Lucile Packard Foundation. We thank M. Z. Hasan for useful discussions. CF thanks J. Liu for helpful discussions and specially thanks T. Mao for his help with the Appendix.
\end{acknowledgments}
%

\begin{appendix}
\section{The equivalence between the two sufficient and necessary conditions of the $\Theta_S$-symmetry}
\label{apn}
In the text, we derived two conditions for the AFM to have $\Theta_S$-symmetry from different physical perspectives. First, $\Theta_S$-symmetry is equivalent to the existence of a lattice vector $\mathbf{D}$, the translation by which inverts all spins. From considering AFM as perturbation, however, $\Theta_S$-symmetry means that for any given set of $m$ integers $z_{1,2,...,m}$, if $\sum_iz_i\bQ_i=\mathbf{G}$, then $\sum_iz_i=even$. These two conditions are asserted to be mathematically equivalent to each other; but the equivalence is not made transparent in the text and is proved in this Appendix. This Appendix is purely mathematical and may be skipped unless one needs to be convinced of the assertion by a formal proof.

First we formulate the statement in a more mathematical fashion. Suppose the commensurate AFM has $m$ propagation vectors, each having $d$ components (in a $d$-dimensional space). We can put these vectors into a $d\times{m}$ rational matrix, where $q_{ij}=\mathbf{a}_i\cdot\bQ_j/2\pi$ is the $i$-th coefficient of $\bQ_j$ expanded in terms of $\mathbf{b}_i$. The lattice vector $\mathbf{D}$ can be expanded in $\mathbf{a}_i$'s, $\mathbf{D}=\sum_{i=1,...,d}D_i\mathbf{a}_i$. In these symbols, the two conditions can be put into the following two statements:\\
For a given matrix $q$,\\
$S_A$: $\exists$ $D=(D_1,...,D_d)\in{Z^d}$, s.t. $2D^Tq\in(2Z+1)^m$;\\
$S_B$: $\forall$ $z\in{Z^m}$ s.t. $qz\in{Z}^d$, there is $\sum_{i=1,...,m}z_i=even$.\\
To understand why the satisfying $S_A$ is equivalent to the existence of $\mathbf{D}$ inverting all spins, simply notice that if $S_A$ is true, there is $\mathbf{D}\cdot\bQ_j=odd*\pi$ for any $j$.

We will follow the steps, detailed in three subsections, to prove that the statements $S_A$ and $S_B$ are equivalent:\\
1. Define a set of linear transformations on any given $q$, $q'=RqC$, where $R$ is an integer $d\times{d}$ matrix and $C$ a $m\times{m}$ matrix with integer entries. Then we prove that if $S_{A,B}(q)=1$, then $S_{A,B}(q')=1$ and vice versa. (Here by $S_{A,B}(q)=0/1$, we mean that $S_{A,B}$ is false/true for matrix $q$.) In other words, $S_{A,B}$ are invariant under these transformations.
2. Prove that using the above transformations, any given $q$ can be transformed into one of the basic forms, to be defined later.
3. Prove that for any matrix $q_0$ of a basic form, $S_{A,B}(q_0)=1$ or $S_{A,B}(q_0)=0$.
There are five basic forms: $B_1=0_{d\times{m}}$; $B_2$ satisfies $(B_2)_1j=p_{1j}/2^k_{1j}$ where $p_{1j}\neq0$ and $k^{1j}=k_{1j'}$ for any $j,j'$; $B_3$'s has at least one column that is zero; $B'_1$ is a diagonal matrix with first $i_c$ elements being in the form $q_{ii}=p_{ii}/2^{k_{ii}}$ where $p_{ii}\neq0$, and all the other elements are zero; $B'_2$ is a block-diagonal matrix with the upper block a diagonal matrix of dimension $i_f-1$ satisfying $q_{ii}=p_{ii}/2^{k_{ii}}$ where $p_{ii}\neq0$ and the lower block in the form of $B_2$.

\subsection{Linear transformations that leave $S_{A,B}$ invariant}

Below we define seven transformations on $q$ that leave $S_{A,B}(q)$ invariant:\\
(i) interchange two rows in $q$;\\
(ii) interchange two columns in $q$;\\
(iii) subtract an integer $n_{ij}$ from $q_{ij}$;\\
(iv) multiply the $i\neq1$-th row by an integer and add it to the first row;\\
(v) multiply the first row by an odd integer;\\
(vi) multiply the first column by an integer $c_i$ satisfying $\sum_{i=1,...,m}c_i=even$ and $c_1=0$, and add all resulted columns to the first column; and\\
(vii) multiply the first column by an odd integer.

By using (i,ii), we can easily extend the definitions of (iv,v,vi,vii) to\\
(iv') multiply the $i\neq{j}$-th row by an integer and add it to the $j$-th row;\\
(v') multiply any row by an odd integer;\\
(vi') multiply the $i$-th column by an integer $c_i$ satisfying $\sum_{i=1,...,m}c_i=even$ and $c_j=0$, and add all resulted columns to the $j$-th column; and\\
(vii') multiply the $j$-th column by an odd integer.

Except for the first two transformations, which mean a permutation of the spatial dimensions and a permutation of the propagation vectors, the physical meaning of the other transformations is not obvious. But the physical meaning is not our concern in the proof, and we only need that under any of these transformations, $S_{A,B}(q)$ is invariant for any $q$. The proofs of the invariance are straightforward for all transformations, yet we prove the invariance under (vi) and (vii) explicitly here.

The transformation in (vi) is expressed as $q'=qC$, where $C_{ij}=\delta_{ij}+r_i\delta_{i1}$ satisfies $\sum{r}_i=even$ and $r_1=0$. First we prove that $S_A(q)=1\rightarrow{S}_A(q')=1$. From $(2D^Tq)_i\in{odd}$, we have $\sum_j(2D^Tq)_jC_{ji}=(2D^Tq)_i$ for $i\neq1$ and $\sum_j(2D^Tq)_jC_{j1}=(2D^Tq)_1+\sum_ir_i(2D^Tq)_i\in{odd}$, i.e., $(2D^Tq')_i\in{odd}$. Then we prove that $S_B(q)=1\rightarrow{S}_B(q')=1$ by exclusion. If $S_B(q')=0$, there is $z'$ such that $q'z'=qCz'\in{Z}^d$ and $\sum{z'}_i\in{odd}$. Defining $z=Cz'$, we have $qz\in{Z}^d$ and $\sum(Cz')_i=\sum{z'}_i+\sum{r}_iz'_i\in{odd}$, so $S_B(q)=0$. Third, we notice that $C^{-1}$ can be obtained by changing $r_i\rightarrow-r_i$, so $C^{-1}$ also belongs to this class of transformations. Therefore, $S_{A,B}(q')=1\rightarrow{S}_{A,B}(q)=1$ follows automatically.

The transformation in (vii) is expressed as $q'=qC$, where $C_{ij}=\delta_{ij}$ for $i\neq1$ and $C_{11}=r\in{odd}$. First we prove $S_A(q)=1\rightarrow{S}_A(q')=1$. Since $(2D^Tq)_i\in{odd}$, and $(2D^Tq')_i=(2D^TqC)_i=(2D^Tq)_i$ for $i\neq1$ and $(2D^Tq')_1=(2D^TqC)_1=r(2D^Tq)_1$, we have $(2D^Tq')_i\in{odd}$. Second we prove $S_A(q')=1\rightarrow{S}_A(q)=1$. Since $(2D^Tq')_i\in{odd}$, we have $[2(rD)^Tq']_i\in{r}*odd$; then there is $[2(rD)^Tq]_i=[2(rD)^Tq'C^{-1}]_i\in{}r*odd*C_{ii}^{-1}\in{odd}$. Third, we prove $S_B(q)=1\rightarrow{S}_B(q')=1$ by exclusion. If $S_B(q')=0$, then there is $z'$ such that $2q'z'=2qCz'\in{Z}^d$ and $\sum{z'_i}\in{}odd$, so $\sum(Cz')_i\in{odd}$, and $z=Cz'$ is an integer vector that satisfies $2qz\in{Z}^d$ and $\sum{z}_i\in{}odd$. Finally we prove $S_B(q')=1\rightarrow{S}_B(q)=1$ by exclusion. If $S_B(q)=0$, then there is $z$ such that $2qz\in{Z}^d$ and $\sum{z}_i\in{odd}$. From $2qz\in{Z}^d$ we have $2qC(C^{-1}rz)\in{Z}^d$ and $\sum(C^{-1}rz)_i\in{odd}$. So we have $z'=C^{-1}rz\in{Z}^m$ such that $2q'z'\in{Z}^d$ and $\sum{z'}_i\in{odd}$. We have shown that under linear transformation described in (vii), $S_{A,B}(q)$ is invariant.

As a final remark, the invariant transformations introduced here do \emph{not} form a group, because the inverse of transformations-(v,vii) involves dividing a row or column by some integer, which do \emph{not} leave $S_{A,B}$ invariant in general.

\subsection{Reduction to basic forms}

Here we show that using the above transformations, every $q$-matrix can be reduced to one of the `basic forms'.

Step one: First we stress that for commensurate AFM, every element $q_{ij}$ must be a rational number. If any $|q_{ij}|\ge1$, we can use transformation-(iii) to send it to a value between $-1$ and $+1$. This is the first step of the reduction, after which every $q_{ij}$ can be expressed as $p_{ij}/s_{ij}$ with $s_{ij}>p_{ij}$. If all elements are zero after the step, we have reached a basic form $B_1=0_{d\times{m}}$.

Step two: If this is not the case, then there is at least one $q_{ij}\neq0$ remaining. (From this point we no longer differentiate $q$ and $q'$, as we have shown that $S_{A,B}$ is invariant under the transformations listed.) Note that every $s_{ij}$ can be written as a product of $2^{k_{ij}}$ and some odd integers, where $k_{ij}\ge0$. Using transformations-(v',vii') we can eliminate all the odd integers from $s_{ij}$. After this we again use transformation-(iii) to eliminate all integer entries. ($q_{ij}$'s whose $s_{ij}$ only has odd integers are put to zero.) Perform the procedure for each $q_{ij}$ whose $s_{ij}$ has an odd factor. This is the second step of reduction. The reduction stops if after this step $q=B_1$.

Step three: If after the second step $q$ is still nonzero, we have $s_{ij}=2^{k_{ij}}$ for every nonzero element of $q_{ij}$. Find the largest $k_{ij}$ and send the corresponding element $q_{ij}$ to $q_{11}$ using transformations-(i,ii). In the next step, we first use transformation-(iii,iv') to eliminate all the other elements in the first column. This can be done as follows: (1) for any $q_{i1}=p_{i1}/2^{k_{i1}}\neq0$, multiply the first row by $2^{k_{11}-k_{i1}}$ and add it to the $i$-th row; (2) now we have $q'_{1i}=(q_{11}+q_{i1})/2^{k_{i1}}$, so one can cancel at least a factor of 2 in the numerator and denominator of $q_{i1}$, after which the new $s'_{i1}=2^{k'_{i1}}$ where $k'_{i1}<k_{i1}$ and (3) repeat (1,2) until the $q_{i1}$ becomes an integer, then use transformation-(iii) to eliminate it. Next we try to eliminate $q_{1j\neq1}$ in the first row. If there is any $q_{1j_1\neq1}=0$, then we can use this element together with $q_{11}$ to eliminate all the other elements in the first row through transformation-(vi'). Here the elimination process is almost the same as the one eliminating $q_{1i\neq0}$ (after replacing rows by columns) with one catch: when $k_{1j}=k_{11}$, $2^{k_{11}-k_{1j}}=1\in{odd}$ so subtracting the first column from the $j$-th column is \emph{not} one of the invariant transformations, but since there is some $q_{1j_1}=0$, we can subtract the first row \emph{and} the $j_1$-th row from the $j$-th row. If $q_{1j}\neq0$ for any $j$, but there is some $k_{1j_2}<k_{11}$, we can use the first row to eliminate $q_{1j_2}$, then use the first and the $j_2$-th row to eliminate all the other elements in the first row. This is the third step in the reduction process. The elimination of the first row cannot proceed if and only if $q_{1j}\neq0$ and $k_{1j}=k_{11}$, that is, when all elements in the first row are nonzero and have the \emph{same} denominator. A $q$-matrix in this form is defined to be of a basic type called $B_2$.

Step four: Suppose after the previous step $q\notin{B}_2$, then we have $q_{1j}=q_{i1}=0$ for $i,j\neq1$ and $q_{11}\neq0$. The $q$-matrix is now block-diagonalized into two blocks, the upper left block having only one element $q^u=q_{11}$, and the lower right block containing the rest $(q^l)_{ij}=q_{i+1j+1}$. For $q^l$, we repeat all the above three steps, with three possible outcomes: $q_l=B_1$, $q_l\in{B}_2$ or $q^l$ is block-diagonalized with its first row and column eliminated except $q^l_{11}$. We define $B'_1$ as the set of matrices having $q_{ii}\neq0$ for $i\le{i_c}>0$ where $0<i_c<d$ and all other elements zero; and define $B'_2$ as the set of block-diagonalized matrices whose upper blocks are diagonal matrices of dimension $i_f-1>0$ and lower blocks are of basic form $B_2$. We see that if we repeat the reduction process finite times, we have the following three possibilities: $q$ is reduced to basic form $B'_1$; q is reduced to basic form $B'_2$; when $d<m$, $q_{ii}\neq0$ for $i\le{d}$ and all other elements are zero, this is of basic form $B_3$ because there is at least one zero column.

We have shown that under a series of the invariant transformations introduced, any $q$-matrix is reduced to one of the five basic forms: $B_1$, $B_2$, $B'_1$, $B'_2$ and $B_3$.

\subsection{Proving $S_A=S_B$ for all basic forms}

In the part, we finish the proof by showing that for any $q$-matrix in a basic form, denoted by $q_0$, $S_A(q_0)=S_B(q_0)$.

If $q_0=B_1$, it is trivial to show that $S_A(q_0)=S_B(q_0)=0$, i.e., both statements are false. Physically, this basic form corresponds to an FM state, which of course does not exhibit $\Theta_S$-symmetry.

If $q_0\in{B}'_1$, we show that $S_A(q_0)=S_B(q_0)=0$, i.e., both statements are false. To see this, simply notice that in $q_0$ there is at least one zero column, namely the $j_0$-th column, therefore $(2D^Tq_0)_{j_0}=0\notin{odd}$ for any $D\in{Z}^d$ and $S_A(q_0)=0$. On the other hand, define a vector $z_0\in{Z}^m$ such that $(z_0)_j=\delta_{jj_0}$, then we have $q_0z_0=0\in{Z}^d$ and $\sum({z}_0)_j=1\in{odd}$, so $S_B(q_0)=0$. Similarly, we can show that for $q_0\in{B}_3$, we also have $S_A(q_0)=S_B(q_0)=0$. Physically, this basic form corresponds to ferrimagnetism, which has modulating local magnetization with a nonzero average.

If $q_0\in{B}'_2$, we show that $S_A(q_0)=S_B(q_0)=1$. We choose $D\in{Z}^d$: $D_i=2^{k_{ii}-1}$ for $i\le{i_f}$ and $D_i=0$ for $i>i_f$, and we have
\bea\label{eq:temp12}
2D^Tq_0=(p_{11},p_{22},...,p_{i_fi_f},p_{i_fi_f+1},...,p_{i_fm}).
\eea
Since all $p_{ij}$'s appearing in Eq.(\ref{eq:temp12}) are odd, $S_A(q_0)=1$. For any $z\in{Z}^m$, if $q_0z\in{Z}^d$, there must be $z_i=2^{k_{ii}}$ for $i<i_f$, and $\sum_{j=i_f,...,m}z_jp_{i_fj}\in{even}$, from which we know $\sum_{i=i_f,...,m}z_i$ must be even, hence $\sum{z}_i\in{even}$. The proof for $S_A=S_B$ is very similar if $q_0\in{B}_2$, because $B_2$ can be seen as a special case of $B'_2$ with $i_f=m$.

We have proved that for $q_0$ in any basic form, $S_A(q_0)=S_B(q_0)$. Combining the results from the previous two subsections, $S_A=S_B$ for any $q$-matrix.

Before closing the Appendix, we gives a few examples how we reduce a commensurate AFM, i.e., a matrix $q$, to its corresponding basic form $q_0$. We assume a 2D system in these examples. For the first example, consider an AFM with one propagation vector $\bQ=(\mathbf{b}_1+\mathbf{b}_2)/3$, so $q=(1/3,1/3)^T$. By transformation-(vii), we have $q\rightarrow(1,1)^T$ and by transformation-(iii), we have $q_0=(0,0)^T\in{B_1}$. In the second example, consider an AFM with two propagation vectors $\bQ_1=\mathbf{b}_1/24+\mathbf{b}_2/4$ and $\bQ_2=3\mathbf{b}_1/24+5\mathbf{b}_2/4$, i.e., $q=\left(\begin{matrix} 
      1/24 & 3/24 \\
      1/4 & 5/4 \\
   \end{matrix}\right)$. First use transformation-(v) to obtain $q\rightarrow\left(\begin{matrix} 
      1/8 & 3/8 \\
      1/4 & 5/4 \\
   \end{matrix}\right)$, then multiply the first row by $-2$ and add it to the second row (transformation-(iv')), which becomes $q_0=\left(\begin{matrix} 
      1/8 & 3/8 \\
      0 & 1/2 \\
   \end{matrix}\right)\in{B}_2$. In the third example, consider an AFM with three propagation vectors $\bQ_1=\mathbf{b}_1/4$, $\bQ_2=\mathbf{b}_2/2$ and $\bQ_3=\mathbf{b}_1/2$, i.e., $q=\left(\begin{matrix} 
      1/4 & 0 & 1/2\\
      0 & 1/2 & 0\\
   \end{matrix}\right)$. To simplify, just use transformation-(vi'), multiplying the first column by $-2$ and add it to the third column, resulting in $q_0=\left(\begin{matrix} 
      1/4 & 0 & 0\\
      0 & 1/2 & 0\\
   \end{matrix}\right)\in{B}_3$. In the next example, $\bQ_1=\mathbf{b}_1/4$ and $\bQ_2=\mathbf{b}_2/3$, i.e., $q=\left(\begin{matrix} 
      1/4 & 0 \\
      0 & 1/3 \\
   \end{matrix}\right)$, which after applying transformation-(vii') followed by transformation-(iii) becomes $q_0=\left(\begin{matrix} 
      1/4 & 0 \\
      0 & 0 \\
   \end{matrix}\right)\in{B'}_1$. For the last example, consider an AFM with $\bQ_1=\mathbf{b}_2/6$ and $\bQ_2=\mathbf{b}_1/2$, i.e., $q=\left(\begin{matrix} 
      0 & 1/2 \\
      1/6 & 0 \\
   \end{matrix}\right)$. First we use transformation-(ii) to interchanged the two columns, and then use transformation-(v') to obtain $q_0=\left(\begin{matrix} 
      1/2 & 0 \\
      0 & 1/2 \\
   \end{matrix}\right)\in{B}'_2$.
   
\end{appendix}

\end{document}